

\input vanilla.sty
\magnification 1200 \hsize=15.2truecm \advance\voffset by 1truecm
\advance\hoffset by .2truecm
\parskip 0pt plus 1pt
\overfullrule=0pt \advance\baselineskip by 1pt

\font\sixrm=cmr6        \font\eightrm=cmr8 \font\gross=cmcsc10
      \font\caps=cmbx8

\nopagenumbers \TagsOnRight


\def\sy{system}                       
\def\qu{quantum}                      \def\mec{mechanic}
\def\qm{\qu\ \mec{}s}                 
\def\ty{theory}                       \def\qt{\qu\ \ty}
\def\me{meas\-ure}                    \def\mt{\me{}ment}

\def\pov{{\gross pov} meas\-ure}      \def\pv{{\gross pv} \me}
\def\pvo{sharp \ob}           \def\povo{unsharp \ob}
     \def\stf{state transformer}

\def\op{operator}                    
\def\ob{observable}                  \def\obs{observables}
\def\sad{self-adjoint}               \def\sop{\sad\ \op}

                    \def\mom{momentum}
\def\cpl{complementar}                

\def\pee#1{{P[#1]}}           
\def\bo#1{{\bold #1}}                    

\def\no#1{\left\|#1\right\|}             
\def\ip#1#2{\left\langle\,#1\,|\,#2\,\right\rangle} 
\def\kb#1#2{|#1\,\rangle\langle\,#2|}    
\def\ket#1{\mid#1\rangle}                
\def\tr#1{\text{tr}\bigl[#1\bigr]}       
         
    \def\ts12{{\textstyle{\frac 12}}}

\def\Rea{{\bold R}}                     \def\eps{\varepsilon}
                         
\def\fii{\varphi}

\def\hi{{\Cal H}}                     
\def\lh{\Cal L(\Cal H)}

\def\of{(\Omega ,\Cal F)}       \def\rb{\bigl(\Rea ,\Cal B(\Rea )\bigr)}


\def\s{$\Cal S$}

                 \def\12pi{\frac 1{2\pi}}


\overfullrule 0pt
\nopagenumbers

\advance\voffset by -1truecm

\advance\vsize by .5cm


\parskip 0pt plus 1pt

\font\gross=cmcsc10
\font\Titel=cmb10 scaled \magstep 2
 1
\font\eightrm=cmr8

\def\R{\hbox{I\kern -0.2em R}}
\def\1{\hbox{1\kern -0.3em I}}

\def\cpl{complementar}

\headline{\centerline{\sixrm Complementarity of Quantum Observables
}\hfill \folio}

\topinsert \centerline{\sixrm Rivista del Nuovo Cimento 18(4) (1995)}
\endinsert

\title{\Titel The Complementarity of Quantum Observables:}
\\
{\Titel Theory and Experiments.}
\endtitle

\vskip .7truecm
\centerline{{\gross Paul Busch}}
\vskip -3pt
\centerline{\eightrm Department of Physics, Harvard University, Cambridge,
MA 02138, USA}
\vskip -3pt
\centerline{\eightrm E-mail: pbusch\@physics.harvard.edu}
\centerline{\gross Pekka J.\ Lahti}
\vskip -3pt
\centerline{\eightrm Department of Physics, University of Turku, 20500 Turku,
Finland}
\vskip -3pt
\centerline{\eightrm E-mail: pekka.lahti\@utu.fi}
\vskip 1truecm
\centerline{\gross Abstract}
\vskip 3pt
\noindent
Three notions of complementarity -- operational, probabilistic, and
value complementarity -- are reanalysed with respect to  the question
of joint \mt{}s and compared with reference to
some  examples of canonically conjugate observables. It is shown that
the joint measurability of noncommuting \ob{}s is a consequence of the
quantum formalism if unsharp observables are taken into account;
a fact not in conflict with the idea of complementarity, which, in
its strongest version, was originally formulated only for sharp observables.
As an illustration of the general theory,
the wave-particle duality of photons is analysed in
terms of complementary path and interference observables and their unsharp
joint measurability.

\vskip 5pt

\noindent {\bf PACS numbers:} 03.65.Bz; 42.50.Wm


\bigskip

{\bf Contents:}

1. Introduction

2. The framework

3. Coexistence

4. Complementarity

5. Examples

\qquad a) position-momentum pairs

\qquad b) components of spin

\qquad c) spin and spin phase

\qquad d) number and phase

6. Photon split-beam experiment

7. Photon interference with a {\gross qnd} path determination

8. Conclusions

References and footnotes

\bigskip


\subheading{1. Introduction}
Recent advances in experimental quantum physics have made it possible to
perform joint measurements of complementary observables of individual
quantum objects, like neutrons or photons.
Despite its apparently contradictory nature, this statement is in full
harmony with the quantum theory,
the theory of complementarity${}^1$, being, in fact,
anticipated in the notion of coexistence.

The concept of complementarity has both probabilistic and operational contents.
When these two aspects are clearly distinguished it becomes evident that
probabilistically complementary observables may also be coexistent so that
they can be measured together. This is exactly the situation encountered in
recent experiments or in proposals for such experiments.

In this work we set out to reanalyse the
notion of complementarity, emphasising from the outset
the distinction between  its probabilistic and operational versions.
We show their interrelations and study their measurement-theoretical
implications.
Various  canonically conjugate pairs of observables, including also
some less known ones,  are revisited and shown to be complementary.
The general results are put to work in an
analysis of new experiments displaying the wave-particle duality for
single photons.

The {\it which path} experiments for photons and other quantum objects
have remained an issue of intensive experimental and theoretical
investigations throughout the history of \qm. The famous two-slit
arrangement, well-known as a source of interference phenomena in
classical light optics, was quickly recognised as an excellent illustration
of the nonobjectivity of quantum observables of individual systems. The
more recent quantum optical split-beam analogue provided by the Mach-Zehnder
interferometer offered the possibility of actually realising the
thought experiments invented by Bohr and Einstein in their attempts
to demonstrate the idea of complementarity or to circumvent the
\mt\ limitations due to the uncertainty relations. The wave-particle duality
has  been strikingly confirmed on the single-photon level
in a series of recent experiments.$^2$
These rather new experimental achievements -- the possibility of investigating
single quantum objects -- were accompanied on the theoretical side
by an elaboration and justification of an individual interpretation of
quantum mechanics capable of appropriately dealing with such experiments.
In such a framework we shall rederive the
mutual exclusiveness of the particle and the wave behaviour,
reflected in  the two options of path determination and
interference measurements. Moreover, we show that both aspects can be
reconciled with
each other where one does not require absolute certainty with respect to the
path nor optimal interference contrast.

The mathematical description of these experiments is sufficiently simple
as to allow for an exhaustive account of the physical situation. In fact, it
turns out that there is not just one description but a variety of them,
each yielding the same experimental figures but nevertheless leading to
totally different mathematical representations
and physical interpretations. In particular,  we encounter
illustrations of an instrumentalistic account, a phenomenological
description  and a realistic picture of physical
experiments.  In the first case one is only concerned with computing the
counting frequencies, treating the whole experimental set-up as a ``black
box" with a variety of control parameters. The second type of account
acknowledges that there is an input system that influences the black
box, the meas\-uring apparatus,
and one may interpret the counting statistics with respect
to this input system. In both cases varying the control parameters amounts to
specifying another measurement.  It is only in the third, realistic, account
that the mathematical language utilised matches the wordings used by the
experimenters in devising the set-up, carrying out the preparations and
measurements, and interpreting the outcomes in terms of the prepared
system. In effect,  the other two approaches must also base their way of
computing on a certain interpretation; they do introduce
a splitting of the whole process into an observing and an observed part
(measurement performed after a preparation); but the {\it cut} is placed at
different locations, and this decision determines the ensuing mathematical
picture.  In addition, different degrees of reality are ascribed to the
phenomena, ranging from  mere measurement outcomes over highly
contextual entities
to something that may be considered as a kind of quantum object.

\subheading{2. The framework}
In order to allow for an appropriately rigorous exposition,
we recall briefly the relevant features of quantum mechanics together with
its measurement theory.
In the  Hilbert space formulation of quantum mechanics
the description of a physical system \s\ is based on a (complex,
separable) Hilbert space $\hi$, with the inner product $\ip{\cdot}{\cdot}$.
We let $\lh$ denote the set of bounded operators on $\hi$.

The basic concepts of {\it observables} and {\it states} will be adopted as
dual pairs; observables being associated (and identified) with
normalised
positive operator valued ({\gross pov}) measures
$E:\Cal F\to\lh$ on some
measurable spaces  $\of$ (the value spaces of the observables),
whereas states are represented  as (and
identified with) positive bounded linear operators $T$ on $\hi$ of trace one.
The value space $\of$ of an observable  is here taken to be the real Borel
space $\rb$, a subspace of it, or a Cartesian product of such spaces.

The familiar concept of an observable as a self-adjoint operator is contained
in the above formulation as a special case. Indeed
if $E$ is a projection valued ({\gross pv}) \me\
[i.e., $E(X) = E(X)^2$ for all
$X\in\Cal F$] and if  its value space is  $\rb$,
then $E$ can be identified with a unique self-adjoint
operator $A$ acting in $\hi$. In this case we write $E^A$ to indicate that
this measure is the spectral measure of $A$.
Observables which are given as \pv{}s are called {\it sharp} observables,
others will be referred to as {\it unsharp} observables$.^3$
A particular class of states are the vector states,
the one-dimensional projections
generated by the unit vectors $\fii$ of $\hi$.
We denote them as $\pee\fii$
or $\kb\fii\fii$, where $\pee\fii(\psi) := \ip{\fii}{\psi}\fii$.

An immediate advantage of this formulation  is
that the measurement outcome probabilities, which form the empirical basis of
quantum mechanics, are now directly available.
Indeed any pair of an observable $E$ and a state $T$ defines a probability
measure $p^E_T$,
$$
p^E_T(X)\ :=\ \tr{TE(X)},\tag 1
$$
for which the {\it minimal interpretation} is adopted:
the number $p^E_T(X)$ is the probability that a measurement of the
observable $E$ leads to a result in the set $X$ when the system is prepared
in the state $T$. If $T$ is a vector state generated by a  vector $\fii$,
we write $p^E_{\fii}$ instead of $p^E_{\pee\fii}$, and we recall that
$p^E_\fii(X) \ =\ \ip{\fii}{E(X)\fii}$.
If $E\equiv E^A$, we also write $p^A_T$ for $p^{E^A}_T$.

It may be appropriate to remark that the representation of
an observable as a \pov , without restricting it to be a \pv ,
not only  formally exhausts
the probability structure of quantum theory
but is even mandatory from the operational point of view.
We shall have
several occasions to see that the measurement outcome statistics
collected under certain circumstances do define an observable not as a
\pv\ but only as a \pov .

In analysing the measurement-theoretical content of the notion of
complementarity we need to rely on some general results of the theory
of measurement.  Therefore we recall that in order to model a measurement of a
given observable $E$ one usually fixes a measuring apparatus $\Cal A$,
with its Hilbert space $\Cal K$, an initial state $T'$ of the apparatus,
a pointer observable $Z$, and a measurement coupling $V$ (a state
transformation of the compound object-apparatus system).  If $T$ is
the initial state of the object system, then $V(T\otimes T')$ is the
object-apparatus state after the measurement. Denoting the
corresponding reduced state of the  apparatus as $W$, one may
pose the {\it probability reproducibility condition} as the minimal
requirement
for $\Cal K, T', Z$ and $V$ to constitute a
measurement of $E$: for any $X$ and $T$,
 $$
p^E_T(X)\ =\ p^Z_W(\bar X),\tag 2
$$
 where $\bar X$ is the value set of the pointer observable
$Z$ corresponding to the value set $X$ of the measured observable $E$.
It is a basic result of the quantum theory
of measurement that every observable can be measured in the above
sense${}^4$.  Moreover, for any observable $E$ there are
 measurements, for which the
initial state of the apparatus is a vector state, the pointer
observable is a \pv , and the meas\-urement coupling is given by a
unitary mapping.  We shall give several examples of such measurements
subsequently.

There is another important  reading of Eq.\ (2). Assume that we are
given a certain device characterised by a measurement coupling $V$,
a pointer  observable $Z$,
and an initial state $T'$ of the apparatus. Then for each value set $X$
of the pointer observable, the mapping $T\mapsto \tr{V(T\otimes T')\, I
\otimes Z(X)}$ is a positive bounded (by unity) linear
functional on the state space
of the object system. This means that the condition
$$
\tr{TF^{T',Z,V}(X)}\ :=\ \tr{V(T\otimes T')\,
I\otimes Z(X)}\tag 3
$$
defines an observable $F^{T',Z,V}$ of the measured system, with
the same value space as the pointer observable. In order to obtain
an observable
with a given value space $\of$ one needs to introduce an
appropriate pointer function $f:\Omega_Z\to\Omega$ assigning to a given
pointer value the corresponding value of the observable. One then has the
system
observable $F^{T',Z,V,f}$ for which
$$
p^{F^{T',Z,V,f}}_T(X)\ :=\ p^{F^{T',Z,V}}_T\bigl(f^{-1}(X)\bigr)\ =\
p^Z_W\bigl(f^{-1}(X)\bigr)\tag 4
$$
for all initial states $T$ of the object system and for all value sets
$X\in\Cal F$ of the measured observable $F^{T',Z,V,f}$.
The condition (2) for
$\Cal K, T',Z$, and $V$ to constitute a measurement of $E$ is thus
equivalent to
the requirement that the actually measured observable $F^{T',Z,V}$ is
$E$ for some pointer function $f$, that is,
$$
F^{T',Z,V,f} = E.\tag 5
$$
This way of reading the probability reproducibility condition corresponds
to the experimenters' actual practice.

As an illustration of the above ideas consider a toy model of a
measurement of an
observable represented by a discrete self-adjoint operator
$A = \sum a_kP_k$, with the eigenvalues $a_k$ and the eigenprojections
$P_k$.
We aim at measuring $A$ by monitoring the position of a particle (apparatus)
confined to moving on a line; the Hilbert space of the apparatus is
thus $\Cal K = L^2(\Rea,dq)$, and the pointer observable is taken to be
the position $Q_{\Cal A}$ of the apparatus.
An appropriate (unitary) measurement  coupling is then
$U=\exp\left(-i\lambda A\otimes P_{\Cal A}\right)$,
which correlates the
observable to be  measured with shifts in the position  of the apparatus.
Let $\pee\phi$ be the initial apparatus state.
If $\pee\fii$ is an initial object state,
then $\pee{U(\fii\otimes\phi)} =
\pee{\sum P_k\fii\otimes\phi_k}$
is the final object-apparatus state, where $\phi_k(x) =
e^{-i\lambda a_kP_{\Cal A}}\phi (x) = \phi(x-\lambda a_k)$
(in the position representation).
Assuming that the spacing between
the eigenvalues $a_k$ is greater than $\frac\delta\lambda$ and that
$\phi$ is supported in
$\bigl(-\frac\delta{2},\frac\delta{2}\bigr)$, then the $\phi_k$ are
supported in the mutually disjoint sets $I_k=\bigl(\lambda a_k-\frac\delta{2},
\lambda a_k +\frac\delta{2}\bigr)$.
With an appropriate choice of the pointer function $f$ the
condition (2) is always fulfilled, that is,
for each $a_k$ and for all $\fii$,
$$
p^A_\fii(a_k)\ =\ p^{Q_{\Cal A}}_W(f^{-1}(a_k))\  =\
p^{Q_{\Cal A}}_W(I_k),\tag 6
$$
where the final apparatus state takes now the form
$W = \sum p^A_\fii(a_k)\pee{\phi_k}$.
This canonical measurement model, which indeed serves as an
$A$-measurement, has some peculiar features: it is value reproducible,
of the first kind, and repeatable.$^3$

As another example of this type of model, we consider
a case where the actually measured observable is not the intended one.
Assume that we wish to measure (a component of)
the position $Q$ of the
object system with the above method, using the coupling
$U = e^{-i\lambda Q\otimes P_{\Cal A}}$ to correlate the object position $Q$
with the
apparatus  position $Q_{\Cal A}$  serving as
the pointer observable. In the position representations, the
initial object-apparatus wave function
$\Psi(q,x)=\fii(q)\phi(x)$ is
transformed into
$\Psi'(q,x)=\bigl(\text{exp}(-i\lambda Q\otimes P_{\Cal A})\Psi\bigr)(q,x)=
\fii(q)\phi(x-\lambda q)$, which allows one to determine
the actually measured observable:
$$
\aligned
\ip{\Psi'}{I\otimes E^{Q_{\Cal A}}(\lambda X)\Psi'}\ &=\ \int_{\bo R}\, dq
\left|\fii(q)\right|^2
\int_{\bo R}\, dq'\,\lambda\,
\left|\phi\bigl(\lambda(q'-q)\bigr)\right|^2 \,I_X(q')\,dq'\\
&=\ \ip\fii{I_X*f\, (Q)\fii}\ =:\ \ip\fii{E^{Q,f}(X)\fii},
\endaligned\tag 7
$$
 where $I_X*f$ denotes the convolution of the characteristic function
$I_X$ of the set $X$ with the confidence function $f$, which is
determined by the coupling constant $\lambda$ and the initial state of
the apparatus: $f(q)\,=\,\lambda\,\left|\phi(-\lambda q)\right|^2$.
The actually measured observable is therefore an unsharp position
$E^{Q,f}:X \mapsto I_X*f\,(Q)$ and not the sharp position $Q$ (with
the spectral measure $E^Q$). This model was already discussed by von
Neumann${}^5$, though at that time the notion of a \pov\ was not yet
available for a definite conclusion.

Another important concept of measurement theory which we shall
apply subsequently is that of a {\it state transformer}. It describes the
possible state changes of the system under a measurement.
Consider a measurement of an observable $E$ given by
$V$, $T'$, and $Z$. For any $X$, there exists a
(non-normalised) state $\Cal I(X)(T)$ such that the conditional expectation
$\text{tr}\bigl[V(T\otimes T')\, B\otimes Z(\bar X)\bigr]$
for any bounded \sop\ $B$ acting in $\hi$ can be expressed as
$\tr{\Cal I(X)(T)\,B}$. In agreement with Eq.\ (2)
the trace (norm) of this state operator is $\tr{\Cal I(X)(T)} = p^E_T(X)$.
According to its definition, $\Cal I(X)(T)$ is (modulo normalisation)
the state of the system after the
measurement on the condition that the pointer observable $Z$ has the
value $\bar X$ after the measurement. In particular,
$\Cal I(\Omega)(T)$ is the state of the system after the \mt\ on the
plain condition that the measurement has been performed.
By construction, the maps $\Cal I(X):T\mapsto \Cal I(X)(T)$ are linear,
positive transformations
 on the trace class and satisfy $\tr{\Cal I(X)(T)}\le\tr{T}$
for all state operators. Such transformations are called
{\it state transformations}.
Hence the \stf\ $\Cal I$ is a state transformation valued measure
$X\mapsto \Cal I(X)$.
Various properties of measurements are conveniently described in terms of
the associated \stf. For instance, a measurement of an observable $E$
is of {\it the first kind} if it does not change the outcome
statistics, that is,
$\tr{TE(X)} = \tr{\Cal I(\Omega)(T)E(X)}$ for all $X$ and $T$,
and it is {\it repeatable} if its repeated application does not lead to
a new result, that is,
$\tr{\Cal I(X)(T)} = \tr{\Cal I(X)^2(T)}$ for all $X$ and $T$.
In the first of the  two examples above the \stf\ is of the form
$$
\Cal I^A(a_k)(T)\ =\ P_kTP_k,\tag 8
$$
which is clearly repeatable, and thus also of the first kind.
In the second example one gets
$$
\Cal I^{Q,f}(X)(T)\ =\ \int_XA_q\,T\,A_q^*\,dq,\tag 9
$$
with
$$
A_q = \sqrt\lambda\,\phi(-\lambda(Q-q)),
\tag 10
$$
which is still of the first kind but no longer repeatable.
In fact, as is well known, no continuous observable admits a repeatable
measurement$.^4$

\subheading{3. Coexistence}
A key concept in the characterisation of quantum mechanical observables
is that of the coexistence of observables which serves to describe
the possibility of measuring together two or more observables.$^6$
Its rudimentary formal expression is the commutativity of \sad\ \op{}s.
For our study of complementarity it will be useful to
recall briefly both the probabilistic and the measurement-theoretical
aspects of coexistence.

An \ob\ $E:\Cal F\to\lh$ is a representation of a class of \mt\ procedures
in the sense that it associates with any state $T$ the probability
$p^{E}_T(X)$ for the occurrence of an outcome $X\in\Cal F$.  For a pair of
\ob{}s $E_1$, $E_2$ the question may be raised as to whether their outcome
statistics $p^{E_1}_T(X)$ and $p^{E_2}_T(Y)$ can be collected within one
common \mt\ procedure for arbitrary states $T$ and sets $X\in\Cal
F_1$, $Y\in\Cal F_2$. Thus one is asking for the existence of a third
observable whose statistics contain those of $E_1$ and $E_2$.  We say that
observables $E_1$ and $E_2$ are coexistent whenever this is the case. More
explicitly, and with a slight generalisation, a collection of observables
$E_i, i\in\bold I$, is {\it coexistent} if there is an observable $E$ such
that for each $i\in\bold I$ and for each $X\in\Cal F_i$ there is a
$Z\in\Cal F$ such that
 $$
p^{E_i}_T(X)\ =\ p^{E}_T(Z) \tag 11
$$
 for all states $T$. In other words, a set of observables $E_i, i\in\bold
I$, is coexistent if there is an observable $E$ such that the ranges $\Cal
R(E_i)\,:=\, \left\{E_i(X)\,|\,X\in\Cal F_i\right\}$ of all $E_i$ are
contained in that of $E$, that is,
$
\bigcup_{i\in\bold I} \Cal R(E_i)\ \subseteq\ \Cal R(E)$.
Observable $E$ is called a joint \ob\ for the $E_i$.

In concrete applications a joint \ob\ $E$ for a coexistent pair $E_1,\,E_2$
will usually
be constructed on the product space $\Omega:=\Omega_1\times\Omega_2$ of
the two outcome spaces, with $\Cal F$ being some $\sigma$-algebra on
$\Omega$ such that $X\times\Omega_2\in\Cal F$, $\Omega_1\times Y\in\Cal F$
for all $X\in\Cal F_1$, $Y\in\Cal F_2$. Thus, if $\Cal F_i=\Cal B(\Rea)$,
then $\Cal F$ will be conveniently chosen as $\Cal B\left(\Rea^2\right)$.
The observables $E_1,\,E_2$ are then recovered from $E$ as its
marginal observables:
$E_1(X)=E(X\times\Omega_2)$, $E_2(Y)=E(\Omega_1\times Y)$.

In order to \me\ jointly two \ob{}s $E_1$ and $E_2$ of a \sy\ \s\ there must
be a single measurement process which allows one
to collect the measurement outcome statistics of both observables.
In the language of measurement theory this means that there is an
apparatus, initially in a state $T'$, a pointer observable
$Z$, and a measurement coupling $V$ such that for any initial
state $T$ of \s,
$$
\aligned
E_1 = F^{T',Z,V,f_1},\\
E_2 = F^{T',Z,V,f_2},\\
\endaligned\tag 12
$$
where $f_1:\Omega_Z\to\Omega_1$ and $f_2:\Omega_Z\to\Omega_2$
are suitable pointer functions.
This shows immediately
that $E_1$ and $E_2$ are coexistent, being in fact
coarse-grained versions of the actually measured observable $F^{T',Z,V}$.
On the other hand, if two \ob{}s
$E_1$ and $E_2$ are coexistent, with a joint \ob\ $E$,
then clearly any measurement of $E$ is a joint measurement of $E_1$ and
$E_2$ in the sense just described.

There is still another intuitive idea of joint measurability which refers
to the possibility of performing order independent sequential measurements.
Indeed assume that a measurement of  $E_1$, performed on a system initially
in state $T$,  is followed by a measurement of
$E_2$. Then $\tr{\Cal I_1(X)(T)E_2(Y)} = \tr{\Cal I_2(Y)\Cal I_1(X)(T)}$
is the joint probability for obtaining results in $X$ for the $E_1$-\mt\
and in $Y$ for the $E_2$-\mt.
It may happen that such sequential probabilities are order independent,
that is, for all $X, Y$ and $T$,
$\tr{\Cal I_2(Y)\Cal I_1(X)(T)} = \tr{\Cal I_1(X)\Cal I_2(Y)(T)}$.
In that case we say that the involved
sequential measurements of $E_1$ and $E_2$ are
order independent. Any two observables
are coexistent whenever they
admit order independent sequential measurements.$^7$
Yet it
appears that the existence of order independent sequential \mt{}s is not
guaranteed by coexistence.

\subheading{4. Complementarity}
The mutual exclusiveness expressed by the term complementarity
refers to the possibilities of predicting \mt\ outcomes as well as to
the value  determinations.
Both of these aspects were discussed  by  Bohr${}^8$
and  Pauli${}^1$. We shall review below two formalisations, referred to as
(\mt-theoretical) \cpl{}ity and probabilistic \cpl{}ity. The former
implies noncoexistence.
In the case of \pvo{}s the two formulations are
equivalent. However, for unsharp observables
the measurement-theoretical notion of complementarity is stronger than the
probabilistic one. It is due to this fact that
probabilistically complementary observables can be coexistent; and this
explains the sense in which  it is possible to speak of
simultaneous measurements of complementary
observables, like position and momentum, or path and interference \ob{}s.

The predictions of  measurement outcomes for two \ob{}s are mutually
exclusive if probability equal to one for some outcome of one \ob\ entails that
none of the outcomes of the other one can be predicted  with certainty.
Observables $E_1$ and $E_2$ are called
{\it probabilistically complementary} if they share the following property:
$$
\aligned
&\text{if}\ \ p^{E_1}_T(X) = 1,\ \text{then}\ \ 0 < p^{E_2}_T(Y) < 1,\\
&\text{if}\ \ p^{E_2}_T(Y)\, = 1,\ \text{then}\ \ 0 < p^{E_1}_T(X) < 1,
\endaligned\tag 13
$$
for any state $T$  and for all bounded  sets
$X\in\Cal F_1$ and $Y\in\Cal F_2$ for which $E_1(X)\ne I\ne E_2(Y)$.

Assume that for observables $E_1$ and $E_2$ the probabilities
$p^{E_1}_T(X)$ and $p^{E_2}_T(Y)$ both equal one for some state $T$ and
some  sets $X$ and $Y$. This is equivalent to saying that
$E_1(X)\fii =\fii$ and $E_2(Y)\fii = \fii$ for some unit vector $\fii$,
that is, $P[\fii] \leq E_1(X)$ and $P[\fii] \leq E_2(Y)$.$^9$
Hence
in that case $E_1(X)$ and $E_2(Y)$ have a positive lower bound.
If these effects  are projection operators, then
$E_1(X)\fii =\fii$ and $E_2(Y)\fii =\fii$
holds exactly when $\fii\in E_1(X)(\hi)\cap E_2(Y)(\hi)$.
Treating similarly the two other cases excluded by (13),
[i.e., $p^{E_1}_T(X)=1$ and $p^{E_2}_T(Y)=0$, or $p^{E_1}_T(X)=0$ and
$p^{E_2}_T(Y)=1$] one observes,
first of all,
that in the case of \pvo{}s the probabilistic complementarity
of a given pair of  observables is equivalent to the disjointness of their
spectral projections:
$$
\aligned
E_1(X)\land E_2(Y)\ &=\ O,\\
E_1(X)\land E_2(\Omega_2\setminus Y)\ &=\ O,\\
E_1(\Omega_1\setminus X)\land E_2(Y)\ &=\ O,
\endaligned\tag 14
$$
for all bounded  sets $X$ and $Y$ for which $O\ne E_1(X)\ne I$
and $O\ne E_2(Y)\ne I$. The above considerations also show that
for \povo{}s condition (14) always implies
(13), but the reverse implication need not hold.
We take condition (14) as the formal definition of the {\it complementarity}
of observables $E_1$ and $E_2$.$^{10}$
Thus we note that any two
complementary observables are also probabilistically complementary,
but not necessarily vice versa.

Position and momentum form the most prominent pair of \cpl{}y \ob{}s.
Yet, among the unsharp position and momentum pairs, which are all
probabilistically \cpl{}y, there are coexistent pairs, thus breaking
the \cpl{}ity.  We return to this example subsequently.

We mention in passing that in the
literature one still finds  another version of complementarity which we shall
term {\it value \cpl{}ity}$.^{11}$
This refers to the case when certain
predictability of some value of $E_1$ implies that all the values of $E_2$ are
equally likely. This concept is not rigorously applicable in the case of
continuous \ob{}s as they have no proper eigenstates.
Nevertheless it is applied also to position and \mom\ in the intuitive
sense that, for example, a sharp momentum (plane wave) state goes along with a
uniform position distribution. Examples of value \cpl{}y \ob{}s are
(the Cartesian components of the)
position and momentum  of a particle in a trap,
mutually orthogonal spin components of a spin-$\frac 12$ object,
the canonically conjugate spin and spin
phase, or the number and phase \ob .
As  is evident, value complementarity implies
probabilistic complementarity.
On the other hand, the examples to be studied below show that value
complementarity  and \mt-theoretical complementarity are logically
independent concepts.

Let us imagine a pair of observables $E_1$ and $E_2$ which are
probabilistically complementary but
coexistent, so that they can be
measured together. Let $E$ be a joint observable, so that
for any $X\in\Cal F_1$ and $Y\in\Cal F_2$ one has $E_1(X) = E(\bar X)$
and $E_2(Y) = E(\bar Y)$ for some $\bar X\in\Cal F$, $\bar Y\in\Cal F$.
We assume that there exists a repeatable joint measurement of $E_1$ and
$E_2$, with the ensuing  $E$-compatible \stf\ $\Cal I$.
Let $X$ and $Y$ be any two bounded sets for which
$O\ne E_1(X)\ne I$ and
$O\ne E_2(Y)\ne I$, and let $T$ be a state for which $p^{E_2}_T(Y)\ne 0$.
If $T_Y:= \Cal I(\bar Y)(T)/p^{E_2}_T(Y)$,
then the  repeatability implies
$p^{E_2}_{T_Y}(Y) = 1$ as well as
$$
p^{E_1}_{T_Y}(X)\ =\ \tr{\Cal I(\bar X)(T_Y)}\ =\
p^{E_2}_{T}(Y)^{-1}\,\tr{\Cal I(\bar X\cap \bar Y)(T)}.\tag 15
$$
If $\tr{\Cal I(\bar X\cap \bar Y)T} = 0$, then
$p^{E_2}_{T_Y}(Y) = 1$ and
$p^{E_1}_{T_Y}(X) = 0$, which is excluded by the second line of Eq.\ (13).
On the other hand,
if $\tr{\Cal I(\bar X\cap \bar Y)T}\ne 0$, then in the state
$T' := \Cal I(\bar X\cap \bar Y)T/p^E_T(\bar X\cap \bar Y)$ one has
$p^{E_1}_{T'}(X) = 1$ and $p^{E_2}_{T'}(Y) = 1$, which is again excluded
by (13). It follows that  $\Cal I$ cannot be repeatable,
and we have the following result.

\proclaim{A. Proposition}
Probabilistically complementary observables do not admit any
repeatable joint measurements.
\endproclaim

Consider next a pair of \cpl{}y observables $E_1$ and $E_2$.
Conditions (14)   imply that
these observables cannot be coexistent.
Indeed, assuming they were coexistent,
let $E$ be a joint observable.
By the additivity of  measures, and with the above notations,
$E(\bar X\cup \bar Y) + E(\bar X\cap \bar Y) = E(\bar Y) + E(\bar X) =
E_1(X) + E_2(Y)$, for all $X$ and $Y$.
If $X$ and $Y$ are bounded sets for which  $O\ne E_1(X)\ne I$, and
$O\ne E_2(Y)\ne I$,
then, by the first line of Eq.\ (14), $E(\bar X\cap \bar Y) = O$.
Therefore, $E_1(X)= E(\bar X\cup \bar Y)-E_2(Y)\leq I-E_2(Y)$, which
contradicts the second line of (14). Hence $E_1$ and $E_2$
cannot be coexistent, and we have proved the following.

\proclaim{B. Proposition}
Complementary observables do not admit any joint measurements.
Neither do they have any order independent sequential measurements.
\endproclaim

Let $\Cal I_1$ and $\Cal I_2$ be \stf{}s compatible with $E_1$ and $E_2$,
respectively. Assume that for some  sets $X$ and $Y$ there is a
state transformation $\Phi$ such that $\Phi\leq \Cal I_1(X)$ and
$\Phi\leq\Cal I_2(Y)$.$^{12}$
Such a state transformation
is a {\it test} of the effects $E_1(X)$ and $E_2(Y)$. It would allow
one to construct a measurement (\stf) which
provides some probabilistic information on both $E_1(X)$ and $E_2(Y).^{13}$
We say that two \stf{}s are {\it mutually exclusive} if no such
test exists, that is, if
$$
\aligned
\Cal I_1(X)\land\Cal I_2(Y)\ &=\ 0,\\
\Cal I_1(X)\land\Cal I_2(\Omega_2\setminus Y)\ &=\ 0,\\
\Cal I_1(\Omega_1\setminus X)\land\Cal I_2(Y)\ &=\ 0,
\endaligned\tag 16
$$
for all bounded  sets $X$ and $Y$ for which  $O\ne\Cal I_1(X)\ne\Cal
I_1(\Omega_1)$ and
$O\ne\Cal I_2(Y)\ne\Cal I_2(\Omega_2)$.
The following is then a \mt-theoretical characterisation of the
complementarity.

\proclaim{C. Complementarity}
Two observables are complementary if and only if  their
associated \stf{}s are mutually exclusive.
\endproclaim

\noindent
This result sharpens the preceding one.

\proclaim{D. Proposition}
Noncoexistent observables do not admit any joint \mt s.\newline
Complementary observables do not admit any joint tests.
\endproclaim

\noindent
With this result we close our general account of complementarity
and we present  some examples next. Sections 6 and 7  provide
then an analysis of
the wave-particle duality of photons in view of these results.


\subheading{5. Examples}

\noindent
{\bf a) Position - momentum pairs.}
The canonically conjugate position and momentum, $Q$ and $P$,
are the basic observables of any localisable object.
They are Fourier equivalent physical quantities and can be
represented as a Schr\"odin\-ger couple${}^{14}$.
This fundamental connection, which results from the Galilei
covariance of the localisation observable,
is the root of the familiar coupling
properties known for these observables.
In view of their relevance to
the joint measurability  of position and momentum,
we collect some of those features here.

The basic coupling properties are, of course, the {\it commutation relation},
$$
QP - PQ\ =\ iI,\tag 17
$$
which holds on a dense domain, and the {\it uncertainty relation},
$$
\Delta (Q,\fii )\,\Delta (P,\fii )\ \geq\ \frac 12, \tag 18
$$
which holds for  all unit vectors $\fii$.

The Fourier connection $P = U_F^{-1}QU_F$  extends also to the
spectral measures, so that one has $E^P(Y) = U_F^{-1}E^Q(Y)U_F$
for all $Y\in\Cal B(\Rea )$.
Thus, if $E^Q(X)\fii = \fii$ and $E^P(Y)\fii = \fii$ for some
vector $\fii$ and bounded sets $X,Y$, then, using the Schr\"odinger
representation, the
function $\fii$ vanishes (almost everywhere) in $\Rea\setminus X$ and its
Fourier transform $\hat\fii$ is an analytic function that vanishes
(almost everywhere) in $\Rea\setminus Y$.
Such a function vanishes everywhere
according to the identity theorem for analytic functions.
Therefore, the following well-known
relations for the spectral projections of
a Schr\"odinger couple (Q,P) are obtained:
$$
\aligned
E^Q(X)\land E^P(Y)\ &=\ O,\\
E^Q(X)\land E^P(\Rea\setminus Y)\ &=\ O,\\
E^Q(\Rea\setminus X)\land E^P(Y)\ &=\ O,
\endaligned\tag 19
$$
for all bounded $X,Y\in\Cal B(\Rea )$.
These relations show the {\it complementarity} of  $Q$ and $P$
both in the sense of (13) and (14).
It may be of interest to note that $E^Q(\Rea\setminus X)\land
E^P(\Rea\setminus Y)\ne O$, for all bounded $X$ and $Y.^{15}$

The uncertainty relations (18) as well as the
complementarity (19) of $Q$ and $P$ imply their {\it total noncommutativity}:
there are no vectors with respect to which $Q$ and $P$ commute, that is,
$$
\left\{\fii\in\hi\,\big | \, E^Q(X)E^P(Y)\,\fii =
E^P(Y)E^Q(X)\,\fii,\ X,Y\in\Cal B(\Rea)\right\}\ =\ \{0\}.\tag 20
$$
This is obvious from (19) but it follows also from (18)$.^{16}$
Interestingly, in spite of their total noncommutativity,
$Q$ and $P$ do have mutually {\it commuting spectral projections}.
Indeed
$$
E^Q(X)\,E^P(Y)\ =\ E^P(Y)\,E^Q(X)\tag 21
$$
whenever $X$ and $Y$ are periodic (Borel) sets satisfying $X = X +2\pi /a$ and
$Y = Y +a.^{17}$
This is a special case of the fact that $Q$
and $P$ admit {\it coexistent coarse-grainings} $f(Q)$ and $g(P)$ for periodic
functions $f$ and $g$, that is,
$$
f(Q)g(P)\ =\ g(P)f(Q),\tag 22
$$
exactly when
$f$ and $g$ are both periodic functions with minimal periods
$\alpha, \beta$ satisfying $2\pi/\alpha\beta\in\bold Z\setminus\{0\}.^{18}$

Position and momentum are complementary observables.
All of their measurements
are mutually exclusive. Therefore, they cannot be measured or even
tested together.
In spite of this impossibility, Heisenberg$^{19}$ boldly maintained that
the position and momentum of a particle can  be determined simultaneously
provided that the  measuring inaccuracies are in accordance with the
uncertainty relations. Today such joint position-momentum measurements
are indeed reported to be feasible. We complete our discussion of this
example by reviewing the measurement-theo\-ret\-i\-cal
model which anticipated this
possibility.

In Sect.\ 2 we showed that a measurement scheme with the
coupling $U = e^{-i\lambda Q\otimes P_{\Cal A}}$ does define an
unsharp position observable $E^{Q,f}$. Similarly, the coupling
$U = e^{i\mu P\otimes P_{\Cal A}}$ yields, via the position monitoring of
the apparatus, a \mt\ of unsharp momentum $E^{P,g}$, where the confidence
function $g$ depends again on the coupling constant and on the
initial apparatus state.
The question then is whether such unsharp position and momentum
observables could be coexistent. A first positive answer is immediately
obtained by showing that combining any two of such position and momentum
measurements does lead to a measurement scheme which defines
a joint \ob\ for a pair of unsharp position and  momentum observables.
To be explicit, consider a measuring apparatus containing  two initially
independent probe particles 1 and 2 prepared in a state
$\pee{\phi_1\otimes\phi_2}$.
Assume that the object system is coupled to the apparatus via the
interaction
$U = \exp\bigl(-i\lambda Q\otimes P_1\otimes I_2
+ i\mu P\otimes I_1\otimes P_2\bigr)$.
If the positions of the two parts of the apparatus are monitored, that is,
if the pointer observable is taken to be $Q_1\otimes Q_2$, then the
system observable $F\equiv F^{\pee{\phi_1\otimes\phi_2},U,Q_1\otimes P_2}$
which is  measured by this scheme has an  unsharp position and unsharp momentum
as its marginal observables: $F(X\times\Rea) = E^{Q,\bar f}(X)$,
$F(\Rea\times Y) = E^{P,\bar g}(Y)$ for all $X$ and $Y$. It is remarkable that
in this case the confidence functions $\bar f$ and $\bar g$
become automatically coupled. They become
{\it Fourier related}, that is, they are of the form
$\bar f(q) = \ip{q}{T_o|q}$ and $\bar g(p) = \ip{p}{T_o|p}$ for  some
positive operator $T_o$ of trace one.$^{20}$
$T_o$ being a state operator, it follows that the distributions
$\bar f$ and $\bar g$ do fulfil the inequality
$$
\Delta(\bar f)\cdot\Delta(\bar g)\ \geq\  \frac 12.\tag 23
$$
In this way Heisenberg's claim  on the joint
measurability of complementary position and momentum observables is
corroborated both in terms of a precise notion of joint \ob{}s as well as
by means of a quantum mechanical \mt\ model.
There is no need to go into further details of the question posed.
It suffices  to mention that  any pair of unsharp position
and momentum observables $E^{Q,f}$ and $E^{P,g}$ are coexistent whenever
the confidence
functions are Fourier related.$^{21}$.
Being coexistent observables, $E^{Q,f}$ and $F^{P,g}$ cannot be
complementary. This can also be confirmed directly since
for any (Borel) subsets $X$ and $Y$ of $\Rea$
$$
\aligned
&F(X\times Y)\leq F(X\times\Rea) = E^{Q,f}(X),\\
&F(X\times Y)\leq F(\Rea\times Y)\, = E^{P,g}(Y).
\endaligned\tag 24
$$
Assume that for some bounded sets $X,Y$ the probabilities
$\tr{TE^{Q,f}(X)}$ and $\tr{TE^{P,g}(Y)}$
both equal one for some vector state $T=\pee\fii$.
This would imply that
both $\fii$ and $\hat\fii$ have bounded supports, which is impossible
for any vector state $\pee\fii$.
Therefore $E^{Q,f}$ and $E^{P,g}$ are probabilistically
complementary observables, whether coexistent or not.

In view of the experimental advances in the trapping of individual particles
it may be of interest to remark that the canonically conjugate
position and momentum observables of a particle in a box are complementary
even though they are clearly not a Schr\"odinger couple.
Avoiding any technical details, we note only that the Cartesian components
$Q_k, P_k$ of the position and momentum of a particle confined to a cube
(of unit length, say)
are complementary and hence also probabilistically complementary$.^{16}$
Since $P_k$ is now discrete, the value complementarity of the pair
$Q_k,P_k$ is readily verified. Indeed if $\fii_n$ is an eigenvector
of $P_k$, then the conjugate position distribution  is uniform:
$\ip{\fii_n}{E^{Q_k}\big((0,x]\big)\fii_n} = x$ for all $0\leq x\leq 1$.

\subheading{b) Components of spin}
Any spin observable $s_{\hat{\bold a}} := \hat{\bold a}\cdot\sigma$
of a spin-$\frac 12$ object
is  given by its spectral projections, the sharp spin properties
$$
E(\pm\hat{\bold a})\ =\ {\textstyle{\frac 12}}\,
(I\pm\hat{\bold a}\cdot\sigma),\tag 25
$$
where $\sigma = (\sigma_x,\sigma_y,\sigma_z)$ are the Pauli
spin operators and $\hat{\bold a}\in\Rea^3$ is a unit vector.
Such observables are mutually complementary,
and also value complementary whenever they are related to
mutually orthogonal directions.

Rotation covariant families of two-valued unsharp spin observables
are generated by the unsharp spin properties, the operators of the form
$$
F(\pm\bo a)\ =\ \ts12\,(I\pm\bo a\cdot\sigma),\qquad\no{{\bo a}}\leq 1.
    \tag 26
$$
Such \ob{}s arise as smeared versions of sharp spin \ob{}s
$s_{\hat{\bold a}}$, the degree of
smearing being determined by the parameter $\no{{\bo a}}.^{22}$

Pairs of unsharp \ob{}s of the form (26) are probabilistically \cpl{}y
but in general not \cpl{}y. We may derive a simple geometric criterion
for the coexistence of such \ob{}s.
To ensure the coexistence of the
unsharp spin properties $F(\bo a_1)$ and $F(\bo a_2)$,
one needs to find an operator $G=\gamma F(\bo c)$ such that
$$\aligned
O\,\leq\,\gamma F(\bo c),\qquad \gamma &F(\bo c)\leq\,F(\bo a_1),\qquad
\gamma F(\bo c)\,\leq\,F(\bo a_2),\\
F(\bo a_1)&+F(\bo a_2)-\gamma F(\bo c)\,\leq\,I.\\
\endaligned\tag 27
$$
Taking into account that $\alpha F(\bo a)\leq \beta F(\bo b)$ is equivalent
to $\no{{\beta \bo b-\alpha\bo a}}\leq \beta-\alpha$, this \sy\ of
inequalities is equivalent to the following one:
$$\aligned
\no{{\gamma\bo c}}\,\leq\,\gamma,\qquad &\no{{\bo a_1-\gamma\bo c}}
\,\leq\, 1-\gamma,\qquad
\no{{\bo a_2-\gamma\bo c}}\,\leq\, 1-\gamma,\\
&\no{{\bo a_1+\bo a_2-\gamma\bo c}}\,\leq\,\gamma.\\
\endaligned\tag 28
$$
Let us denote by $S(\bo a,r)$ the closed ball with radius $r$ and centre
point $\bo a$. Then (28) can be rewritten as follows:
$$
\gamma\bo c\ \in\ S(\bo a_1,1-\gamma)\cap S(\bo a_2,1-\gamma)\cap
S(\bo a_1+\bo a_2,\gamma)\cap S(\bo o,\gamma).\tag 29
$$
The intersection of the first two balls is nonempty exactly when
$\gamma\leq 1-\frac 12\no{{\bo a_1-\bo a_2}}$, while the intersection of the
last two balls is nonempty if and only if $\gamma\geq\frac 12\no{{\bo
a_1+\bo a_2}}$. Thus the coexistence of $F(\bo a_1)$ and $F(\bo a_2)$
implies the inequality $\frac 12\no{{\bo a_1+\bo a_2}}\leq 1-\frac 12\no{{
\bo a_1-\bo a_2}}$. Conversely, the validity of this relation entails the
existence of some $\gamma$ satisfying the preceding two inequalities.
In turn, these ensure that $\bo c_o:=\frac 12(\bo a_1+\bo a_2)$ is in the
intersection of all four balls. Taking for $\gamma$ the norm of this  vector
and defining $\bo c:=\bo c_o/\no{{\bo c_o}}$, one has satisfied (29).
We have thus established the following result.

\proclaim{E. Proposition} The unsharp  spin
properties $F(\bo a_1)$ and $F(\bo a_2)$
are coexistent if and only if
$$
\no{{\bo a_1+\bo a_2}}\, +\, \no{{\bo a_1-\bo a_2}}\ \leq 2.
\tag 30
$$
\endproclaim

\noindent
If $\no{{\bo a_1}}=1$, say, so that
$F(\bo a_1)$ is a sharp spin property, then (30) is fulfilled exactly when
$\bo a_2=\pm\no{{\bo a_2}}\,\bo a_1$, that is, $F(\bo a_2)$ commutes with
$F(\bo a_1)$. This confirms the general result that the coexistence of
two \ob{}s amounts to their commutativity if one of them is a sharp \ob.
For the coexistence of two noncommuting unsharp \ob{}s a sufficient degree
of unsharpness is required so that the values of $\no{{\bo a_1}},\no{{\bo
a_2}}$
must not be too close to unity.

A joint \ob\ for a pair of coexistent unsharp spin properties
$F(\bo a_1)$, $F(\bo a_2)$
can easily be constructed. For example, the following set of operators will
do:
$$
G_{ik}\ :=\ \alpha_{ik}\,F\left(\textstyle{\frac 1{\alpha_{ik}}\,\frac 12}\,
(\bo a_i+\bo a_k)\right)\tag 31
$$
where $\bo a_i$ is one of $\bo a_1$, $\bo a_{\bar 1}=-\bo a_1$, and
$\bo a_k$ is either $\bo a_2$ or $\bo a_{\bar 2}=-\bo a_2$. The factor
$\alpha_{ik}$ can be taken to be $\frac 12(1+\bo a_i\cdot\bo a_k)$.
It is readily verified that (30) ensures the positivity of all $G_{ik}$
and that, for example,  $F(\bo a_1)=G_{12}+G_{1\bar 2}$.

\subheading{c) Spin and spin phase}
The conjugate pair of a spin  and   spin phase
constitute
another example of probabilistically complementary observables which may or
may not be complementary.
A spin phase  conjugate to $s_3 = \hat e_3\cdot\sigma$, say,
can easily be constructed from the polar decomposition of the raising operator
$s_+ := s_1+ is_2$. The construction applies to any spin $s$, so that
we consider the spin Hilbert space $\hi_s = {\bo C}^{2s+1}$ with
a complete orthonormal set of $s_3$-eigenvectors $\ket m$,
$m = -s, -s+1,\cdots, s-1, s$.
From the  polar decomposition $s_+\ =\ B\,|s_+|$,
with $|s_+|^2 = \sqrt{(s_1-is_2)(s_1+is_2)}$, one easily computes that
the partial isometry $B$ has the structure
$B = \sum_{m=-s}^{s-1}\kb{m+1}{m}$.  This shows that
$B$ is a contraction with the  norm $\| B\| =1$.
Hence there is a unique \pov\ $S$  such that
 $$
B^n\ =\ \int_0^{2\pi}e^{in\alpha}\,S(d\alpha),\tag 32
$$
 for each $n=0,1,2,\ldots$.$^{23}$
We  call $S$ {\it the spin phase}.
The commutation relation $[B,s_3] = B$ implies that
$e^{-i\alpha s_3}\,B\,e^{i\alpha s_3}\ =\ e^{-i\alpha}B$,
$\alpha\in[0,2\pi]$. Due to the uniqueness of the \pov\ $S$, the spin
phase satisfies the {\it covariance condition}
 $$
e^{-i\alpha s_3}\,S(X)\,e^{i\alpha s_3}\ =\ S(X + \alpha),\eqno(33)
$$
modulo $2\pi$ for arbitrary $X\in\Cal B[0,2\pi]$.
The explicit structure of $S$ can easily be obtained.

Let
$\widetilde\hi\,:=\,L^2\bigl([0,2\pi],\frac{d\alpha}{2\pi}\bigr)$ and consider
the following complete orthogonal system of normalised vectors
$\psi_m\in\widetilde\hi$,
$\alpha\mapsto\psi_m(\alpha)=e^{im\alpha}$. The map
 $$
W_s:\hi_s\ \to\ \widetilde\hi ,\quad
   \psi\ \mapsto\ \sum_{m=-s}^s \ip{m}{\psi}\,\psi_m\tag 34
$$
 is an isometry from $\hi_s$ on to the ($2s+1$)-dimensional subspace
$\widetilde\hi_s:=\text{span}\bigl[\psi_m|m=-s,\cdots,s\bigr]$
of $\widetilde\hi$.
In $\widetilde\hi_s$ the operator $B$ is the Neumark
projection of the unitary operator
$
\widetilde B_o = \sum_{m=-\infty}^{\infty}\,\kb{\psi_{m+1}}{\psi_m}$
on $\widetilde\hi$.
Since $(\widetilde B_o\psi)(\alpha) = e^{i\alpha}\psi(\alpha)$,
$\psi\in\widetilde\hi$,
the spectral measure of $\widetilde B_o$ is the canonical one
$X\mapsto E(X)$, with $E(X)\psi = I_X\psi$. Thus one gets:
 $$
 S(X)\ = \sum_{m,n=-s}^s\,\ip{\psi_m}{E(X)\psi_n}\,\kb mn\ =
\sum_{m,n=-s}^s \int_X \,e^{i(n-m)\alpha}\,\kb mn
\,\textstyle{\frac{d\alpha}{2\pi}}.\eqno(35)
$$

It is important to note that $S$ itself is no \pv.
The space $\widetilde\hi_s$ consists of finite linear
combinations of trigonometric functions. The idempotency
$S(X) = S(X)^2$ would demand that $I_X\xi\in\widetilde\hi_s$ whenever
$\xi\in\widetilde\hi_s$. This is impossible since such functions
cannot be represented as finite linear combinations of sine and
cosine functions.

For the same reasons the spin phase cannot be ``localised" since for
any proper subset $X$ of $[0,2\pi]$ (with $0<\int_Xd\alpha<2\pi$) and
all $\xi\in\widetilde\hi_s$ one has
 $$
0\ <\   \ip{\xi}{E(X)\xi}\ <\ 1.\eqno(36)
$$
Thus the pair $(s_3,S)$ is trivially probabilistically
complementary.
It is also value complementary since in any spin eigenstate $\ket m$ the
spin phase is uniformly distributed:
$\ip{m}{S\bigl([0,\alpha]\bigr)m} = \frac {\alpha}{2\pi}$.
On the other hand, while these \ob{}s are noncoexistent, they are not
\cpl{}y: for any spin state $\ket m$  there is
a positive number $\lambda<1$ such that $\lambda\kb mm$ is a lower
bound to $\kb mm$ as well as to any (nonzero) $S(X)$. This follows from a
result$^{24}$ according to which $\lambda\pee\fii$ is a lower bound of
an operator $A$, $O\leq A\leq I$,
 for some positive $\lambda$ exactly when $\fii$ is in the
range of the square root of
$A$; in the present case of $A=S(X)$ this range is the
whole Hilbert space due to the validity of (36) for any state.

\subheading{d) Number and phase}
It may be noted that similar results are obtained for the
conjugate pair of photon number and the phase of a single-mode
electromagnetic field. The only formal difference consists in the fact that
the number \op\ $N=a^*a$ is semibounded. The phase, defined as a \pov\
$M$ via the polar
decomposition of the annihilation \op\ $a = B\sqrt{N}$
analogously to (32), is
shift covariant under the group generated by $N$.$^{25}$ Again the phase is not
localisable so that the pair $(N,M)$ is probabilistically complementary;
moreover it is value \cpl{}y as the phase is uniformly distributed in
any eigenstate of the number.
%

\subheading{6. Photon split-beam experiments}
Using the tools and concepts developed in Sections 2-4,
we  analyse the photon split-beam experiments
performed by a Mach-Zehnder interferometer.
Figure 1 shows the scheme of such a device consisting of
two beam splitters $BS_1$ and $BS_2$, with
transparencies $\eps_1$ and $\eps_2$, reflecting mirrors $M_1$ and
$M_2$, and a phase shifter $PS(\delta)$ allowing for the variation of the
path difference between the two arms of the interferometer.  The detectors
$D_1$ and $D_2$ are assumed to register the number of photons $N_1 = \sum
n_1\kb{n_1}{n_1}$ and $N_2 = \sum n_2\kb{n_2}{n_2}$ emerging from the
second beam splitter, when a photon pulse is impinging on the first one.
We shall take the incoming photon pulse to be a single-mode field
in a state $T$, and we let $N_a = a^*a$ denote its number operator.  The
first beam splitter effects a coupling of this mode with an
idle single-mode field, with $N_b = b^*b$.  In the Schr\"odinger picture
one may identify $N_1$ and $N_2$ with $N_a$ and $N_b$, respectively.
It will be useful to consider
first arbitrary states $T'$ of the $b$-mode, and only later fix
it to be idle, $T' = \kb 00$.

The action of a beam splitter $BS$ with transparency $\eps$
can be described by means of the unitary \op$^{26}$
$$
U_{\alpha}\ =\ \exp(\bar\alpha a\otimes b^*-\alpha a^*\otimes b),\tag 37
$$
with $\alpha = |\alpha|e^{i\vartheta}$, $\cos|\alpha| =\sqrt\eps$.
The phase shifter $PS(\delta)$ acts according to
$$
V_{\delta}\ =\ e^{i\delta N_a}\otimes I.\tag 38
$$
Therefore, if $T\otimes T'$ is the initial state of the two-mode field
entering the interferometer, then the state of the field emerging from
the second beam splitter is
$$
W\ :=\ U_{\beta}V_{\delta}U_{\alpha}\,(T\otimes T')\, U_{\alpha}^*
V_{\delta}^*U_{\beta}^*,\tag 39
$$
where $\alpha = |\alpha|e^{i\vartheta_1}, \cos|\alpha| = \sqrt{\eps_1}$,
and $\beta = |\beta|e^{i\vartheta_2}, \cos|\beta| = \sqrt{\eps_2}$.
The probability of detecting $n_1$ photons in $D_1$ and
$n_2$ photons in $D_2$ is thus
$$
p^{N_1\otimes N_2}_W(n_1,n_2)\ =\ \ip{n_1,n_2}{W|n_1,n_2}.\tag 40
$$

\line{}

\noindent
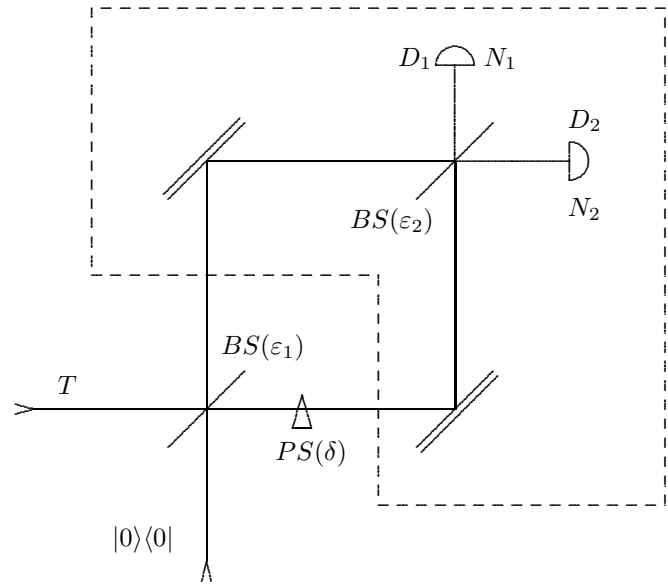
[{\it Figure 1. Scheme of a Mach-Zehnder interferometer.}]

\line{}

\noindent In this reading the measured \ob\ is the two-mode
number observable $N_1\otimes N_2$,
and the \me{}d \sy\ is the output field emerging from the
interferometer. The field's passage through the interferometer is treated
as an indivisible part of the preparation of the phenomenon to be observed,
and no attempt is made at analysing the various stages of this process.  This
view corresponds to a positivistic attitude according
to which \qt\  is merely an instrument for calculating
the statistics of measurement outcomes.
All adjustable variables, the two-mode input state,
the transparencies, and the phase shift, are treated on equal footing as
control parameters of the black box determining the preparation of the
output state $W$, which is subjected to a counting \mt.

The counting probability (40) can equivalently be written as a measurement
outcome probability
with respect to the input state $T\otimes T'$ of the two-mode field,
$$
p^{E}_{T\otimes T'}(n_1,n_2)\ \equiv\ \tr{T\otimes T'\,E(n_1,n_2)}
\ :=\ p^{N_1\otimes N_2}_W(n_1,n_2),\tag 41
$$
the \ob\ being now
$$
E(n_1,n_2)\ :=\ U_{\alpha}^*V_{\delta}^*U_{\beta}^*\bigl(\kb{n_1}{n_1}\otimes
\kb{n_2}{n_2}\bigr)U_{\beta}V_{\delta}U_{\alpha}.\tag 42
$$
Here the interferometer is taken as a part of the measuring device, which now
serves to yield information on the input state. Accordingly, the set of
variables mentioned above is split into two parts, the input state
representing the preparation, and the interferometer parameters
belonging to the \mt. In order to realise wave or particle phenomena,
one must take into account, in this view, all the details of the experiment,
including the measuring system parameters. The `particle' or `wave' cannot
be described as an entity existing independently of the constituting
\mt\ context. In fact the input state alone does not determine whether the
field passing the interferometer behaves like a particle or a wave.

To work out the explicit form of Eq.\ (42), we note first that
$V_\delta U_\alpha=U_{\alpha'}V_\delta$, with $\alpha'=e^{i\delta}\alpha$.
Next, observe that the operators $a\otimes b^*$, $a^*\otimes b$, and
$\frac 12(N_a\otimes I
- I\otimes N_b)$ satisfy the standard commutation relations of
the generators of the group $SU(2)$. Therefore, for any $\alpha$ and $\beta$,
there is a $\gamma$ such that$^{27}$
$$
U_{\alpha}U_{\beta}\ =\ e^{\frac i2f(\alpha,\beta)(N_a\otimes I-I\otimes N_b)}
\,U_{\gamma}.\tag 43
$$
This allows one  to write the  observable $E$  as
$$
E(n_1,n_2)\ =\ E^{\eps}(n_1,n_1) \ :=\ U^*_{\gamma}\,\kb{n_1}{n_1}\otimes
\kb{n_2}{n_2}\,U_{\gamma},\tag 44
$$
for an appropriate $\gamma$. It follows that the
Mach-Zehnder interferometer acts like a single  beam splitter with transparency
$\eps$, where $\sqrt\eps = \cos|\gamma|$, and
$\gamma = \gamma (\alpha,\beta,\delta)$.
The explicit form of $\gamma$, and thus of
the effective transparency $\eps$ can be most easily calculated by
evaluating, for example, the single-photon probability $p^E_{\ket{10}}(1,0)$,
which is done below.

If the state $T'$ of the second mode is kept fixed, one may view the
$a$-mode alone as the input \sy\. This step is necessary if one wants to
represent the idea that a light pulse, or even one photon, coming from
one source is subjected to an interferometric \mt. In that view
the counting statistics (40)
defines an observable $F^{T',\eps}$ of this mode such that for
any $T$ and for all $n_1, n_2$ one has
$$
\tr{T F^{T';\eps}(n_1,n_2)} \
:=\ \tr{T\otimes T' E^{\eps}(n_1,n_2)}.\tag 45
$$
If $T'$ is a vector state, then this observable is just the
Neumark projection of $E^{\eps}$ with the projection $I\otimes T'$:
$$
F^{T';\eps}(n_1,n_2) \ \equiv\ I\otimes T' E^{\eps}(n_1,n_2) I\otimes T'.\tag 46
$$
Finally, taking the second mode to be in the vacuum state,
the explicit form of this observable is obtained by a simple computation:
$$
F^{0;\eps}(n_1,n_2) \ =\
\textstyle{\frac{(n_1+n_2)!}{n_1!n_2!}}\,\eps^{n_1}\,(1-\eps )^{n_2}\,
\kb{n_1+n_2}{n_1+n_2}.\tag 47
$$

The counting statistics of a single mode light field
sent through an interferometer defines thus an observable of this field.
This \ob\ is a \pv\ exactly when
$\eps = 0$, 
or $\eps = 1$. 
Any choice of  the parameter $\eps$ refers to a different
experimental arrangement consisting of
$BS_1$, $BS_2$, and $PS(\delta)$.
As a rule,
they all give rise to different observables $F^{0;\eps}$, which are
thus mutually exclusive in the trivial sense that any choice of
$\eps_1$, $\eps_2$, and $\delta$ excludes another one.
However,  these observables all are mutually
commuting. Nevertheless they
reflect {\it on a phenomenological level} the wave-particle duality of
a single photon, as will become clear below. We call this description
phenomenological as the production of the wave or the particle phenomena is
described in terms of the instrumental tools, without making reference to
the behaviour of the observed entity.

The marginals
$F^{0;\eps}_i$, $i=1,2$, of the observable $F^{0;\eps}$ are found
to be of the form
$$
\aligned
F_1^{0;\eps}(n) \ &=\
\sum_{m=n}^\infty{\textstyle{\binom mn}}\eps^n(1-\eps)^{m-n}\kb mm,\\
F_2^{0;\eps}(n)\ &=\ F_1^{0;1-\eps}(n).
\endaligned\tag 48
$$
They are mutually commuting
observables representing unsharp versions of
the number observable $N_a$.

We consider next the case of the incoming $a$-mode being prepared in a
number state $T = \kb nn$. The probabilities (45) obtain then the simple form
$$
\aligned
p_{\ket n}^{F^{0;\eps}}(n_1,n_2) \
&=\ \ip{n}{F^{0;\eps}(n_1,n_2)|n}\\
&=\ \textstyle{\frac{(n_1+n_2)!}{n_1!n_2!}}\,\eps^{n_1}\,
(1-\eps)^{n_2}\,\delta_{n,n_1+n_2}.
\endaligned\tag 49
$$
From these probabilities the conservation of photon number (energy) in the
interferometer is manifest: $n$ input photons give rise to a total
of $n=n_1+n_2$ counts in the two detectors.

We are now ready to discuss the
wave-particle duality for a single photon  input
$T = \kb 11$. Formula (49) gives
$$
\align
\ip{1}{F^{0;\eps}(1,0)|1} \ &=\ \eps,\tag 50a\\
\ip{1}{F^{0;\eps}(0,1)|1} \ &=\ 1-\eps,\tag 50b\\
\ip{1}{F^{0;\eps}(n_1,n_2)|1} \ &=\  0 \ \ \text{if}\ n_1+n_2\neq 1.
\tag 50c\\
\endalign
$$
The anticoincidence of counts expressed in (50c) is an
indication of the corpuscular nature of photons: one single photon
cannot make two detectors fire.
In this case the dependence of $\eps$ on the parameters of
the interferometer is easily determined:
$$
\eps\ =\
\eps_1\eps_2\, +\, (1-\eps_1)(1-\eps_2)\,+\,
2\sqrt{\eps_1(1-\eps_1)\eps_2(1-\eps_2)}\,
\cos(\vartheta_2-\vartheta_1-\delta).\tag 51
$$

There are three cases of special interest:
$\eps_1$ variable, $\eps_2 = 1$;
$\eps_1 = \eps_2 =\frac 12$; and
$\eps_1=\frac 12$, $\eps_2$ variable.
The first choice gives $\eps = \eps_1$, the second
$\eps = \cos^2\left({\frac 12}(\vartheta_2-\vartheta_1+\delta)\right)$,
and the third $\eps = \frac 12\left[1+2\sqrt{\eps_2(1-\eps_2)}
\cos(\vartheta_2-\vartheta_1-\delta)\right]$ =
$\frac 12\left[(1+2\sqrt{\eps_2(1-\eps_2)}
\cos\delta\right]$, where the last equality is under the assumption that
$\vartheta_1=\vartheta_2$. Some of these cases have been investigated
experimentally, confirming thus the predicted quantum mechanical
probabilities (50).$^{2}$

The experiment with $\eps_2 = 1$ would allow one to decide from a
single count event whether $\eps_1$ was 1 or 0 if one of these values
was given. This is interpreted as the calibration for a path \mt.  If
$\eps_1$ differs from these values, then, of course, the notion of a
path taken by the photon is meaningless. But the very ability of the device
to detect the path (if it was fixed) destroys any interference.  Next, the
statistics obtained in the case $\eps_2 = \frac 12$ reproduce the
expected interference pattern resulting from many runs of this
single-photon experiment. More precisely, the interference disappears if
$\eps_1$ is 0 or 1, which corresponds to the situation where the photon
is forced to take exactly one path. In this sense a precise fixing of the
path again destroys  the interference. Maximal path indeterminacy,
$\eps_1=\frac 12$, gives rise to optimal interference, while there is
no way of getting  any information on the path when $\eps_2=\frac 12$.
In this way we recover the wave-particle duality in Bohr's complementarity
interpretation. There are mutually exclusive options for both, the
{\it preparation}, as well as the {\it registration},
of path or wave behaviour.

In a recent experiment$^{28}$
a modified Mach-Zehnder interferometer, with $\eps_1 = \frac 12$ fixed
and variable $\eps_2$,
was introduced for testing the detection
probability
$$
\frac 12\left[1+2\sqrt{\eps_2(1-\eps_2)}\cos\delta\right].
$$
This experiment
was interpreted as providing simultaneous information on the two complementary
properties of a photon.
Indeed, letting $\eps_2$ vary from $\frac 12$
to 1, one recognises that the interference fades away gradually from
the pattern with  optimal contrast,
$\cos^2\left(\frac 12(\vartheta_2- \vartheta_1+\delta)\right)$,
to no interference at all, $\frac 12$.
The experimentally realised case $\eps_2 = 0.994$
still leads to a recognisable interference pattern
$\bigl(\eps = \frac 12(1+0.154\cos\delta)\bigr)$ even though there is, loosely
speaking, already a high ($84 \%$) confidence on the path of
the photon. In a suitable \me, this situation was characterised by ascribing
$98.2\%$ particle nature and $1.8\%$ wave nature to the photon.
Unfortunately, in this  experiment$^{28}$ the incoming light
pulses originated from a laser so that no
genuine single photon situation was guaranteed. That is, the intensity was
low enough to ensure, with high probability, the presence of only one photon
in the interferometer, but the detection was not sensitive to single counts.

The analysis carried out so far rephrases the common view that the
detection statistics of single-photon Mach-Zehnder interferometry exhibit
both the wave-particle duality as well as the unsharp wave-particle
behaviour for single photons. Note, however, that the language used here
goes beyond the formal description that could be given in terms of the
observables $F^{0;\eps}$.
An  account based solely on the latter is phenomenological
in the sense
that the relevant photon observables $F^{0,\eps}$, which pertain to the
object under investigation, are mutually commutative; hence on the object
level there is no complementarity.  Only the various statistics for single
photon input states display the {\it complementary phenomena} in question.  We
shall now  change  our point of view to show that the same statistics can
also be interpreted on the basis of {\it complementary observables}.

To this end we redefine again the cut to be placed in the experimental
set-up of Figure 1. Instead of taking the
whole interferometer together with the detectors as the registration device,
we consider the first beam splitter $BS_1$ and the phase shifter
$PS(\delta)$ as  parts of the preparation device. The object system
is therefore the two-mode field  prepared in a state
$$
S\ :=\ V_{\delta}U_{\alpha}\,(T\otimes T')\,U_{\alpha}^*V_{\delta}^*.\tag 52
$$
The detection statistics (40) can then be written as
$$
p^{E^{\beta}}_S(n_1,n_2)\ =\ p^{N_1\otimes N_2}_W(n_1,n_2)\tag 53
$$
for the observable $E^{\beta}$,
$$
E^{\beta}(n_1,n_2)\ :=\ U^*_{\beta}\,\kb{n_1}{n_1}\otimes\kb{n_2}{n_2}\,
U_{\beta}.\tag 54
$$

We restrict our attention to  the single photon case,
$T=\kb 11$, $T'=\kb00$,
so that the possible initial states of the two-mode field are
the vector states
$$
\psi_{\alpha,\delta}\ :=\ V_{\delta} U_{\alpha}\ket{10}\
=\ \sqrt{\eps_1}\ket{10}\, +\,
e^{-i(\vartheta_1+\delta)}\sqrt{1-\eps_1}\ket{01},\tag 55
$$
where, for instance, $\ket{10} = \ket 1\otimes\ket 0$. Let $P_{10}$
and $P_{01}$ denote the one dimensional
projections  $\kb 11\otimes\kb 00$ and $\kb 00\otimes\kb 11$ of the two-mode
Fock  space. Then $P_{10}+P_{01}$ projects on to the two dimensional subspace
of the vectors (55) which we take to represent the object system to be
investigated. Due to the number conservation under the unitary map $U_\beta$
the projection operators  (54)
commute with $P_{10}+P_{01}$, so that the following
operators define
a \pv\ on the state space of the object \sy :
$$
F^{1,0;\eps_2}(n_1,n_2)\ :=\
\bigl(P_{10}+P_{01}\bigr)\,E^{\beta}(n_1,n_2)\,\bigl(P_{10}+P_{01}\bigr).
\tag  56
$$
For the states (55) the \ob{}s $E^\beta$ and $F^{1,0;\eps_2}$ have the
same expectations,
$$
\ip{\psi_{\alpha,\delta}}{F^{1,0;\eps_2}(n_1,n_2)\psi_{\alpha,\delta}}
\ =\ \ip{\psi_{\alpha,\delta}}{E^{\beta}(n_1,n_2)\psi_{\alpha,\delta}},
\tag  57
$$
for all $n_1,n_2$ and for each $\alpha$ and $\delta$.
In particular, this means that the observable $F^{1,0;\eps_2}$ is
in fact determined by the detection statistics.

The operators of Eq.\ (56) read as follows:
$$
\align
F^{1,0;\eps_2}(1,0)\ \ \ &=\ E^{\beta}(1,0)\
                          =\ P\left[U^*_{\beta}\ket{10}\right]\tag 58a\\
&=\ P\left[\sqrt{\eps_2}\ket{10}+e^{-i\vartheta_2}\sqrt{1-\eps_2}
\ket{01}\right],\\
F^{1,0;\eps_2}(0,1)\ \ \ &=\ E^{\beta}(0,1)\
                          =\ P\left[U^*_{\beta}\ket{01}\right]\tag 58b\\
&=\ P\left[\sqrt{1-\eps_2}\ket{10}-e^{-i\vartheta_2}\sqrt{\eps_2}
\ket{01}\right],\\
F^{1,0;\eps_2}(n_1,n_2)\ &=\ O\ \ \text{if}\ n_1+n_2\neq 1.\tag 58c
\endalign
$$
Also, for any $\alpha$ and $\delta$
$$
\align
\ip{\psi_{\alpha,\delta}}{F^{1,0;\eps_2}(1,0)\psi_{\alpha,\delta}}\ \  &=\
\eps,\tag 59a\\
\ip{\psi_{\alpha,\delta}}{F^{1,0;\eps_2}(0,1)\psi_{\alpha,\delta}}\ \  &=\
1-\eps,\tag 59b\\
\ip{\psi_{\alpha,\delta}}{F^{1,0;\eps_2}(n_1,n_2)\psi_{\alpha,\delta}}\ &=\
0\ \ \text{if}\ n_1+n_2\neq 1,\tag 59c
\endalign
$$
with $\eps$ given by Eq.\ (51).

On the level of a statistical description the change of  viewpoint
has brought nothing new;
the measurement outcome probabilities (59) are, as they should be, the same
as those of Eq.\ (50), or just the detection
statistics (40). There is, however, an essentially new aspect in the
description. All the photon observables $F^{1,0;\eps_2}$ are mutually
complementary in the strong measurement-theoretical  sense: these
observables cannot be measured or even tested together. In particular,
the observables $F^{1,0;1}$ and $F^{1,0;\frac 12}$
associated with the
extreme choices $\eps_2 = 1$ and $\eps_2 = \frac 12$
are  complementary path and interference observables, 
$$
\align
F^{1,0;1}(1,0)\ &=\ P_{10},\tag 60a\\
F^{1,0;\frac 12}(1,0)\ &=\
P\left[\textstyle{\frac 1{\sqrt 2}}\ket{10}+\textstyle{\frac 1{\sqrt 2}}
e^{-i\vartheta_2}\ket{01}\right].\tag 60b
\endalign
$$
In Ref.\ 28 a \mt\ of $F^{1,0;\eps}$ was interpreted as a
joint unsharp determination  of the complementary path and interference
observables $F^{1,0;1}$ and $F^{1,0;\frac 12}$. While  the \ob\
$F^{1,0;\eps}$ does entail probabilistic information on the two
complementary observables, it is a sharp observable, and cannot therefore
be considered to represent
an unsharp joint \mt\ in the sense of the general point of view
adopted in this work. In the next section a proposal is
(re)analysed which does lead to such a joint \mt, as can be read off from the
ensuing \pov{}s.

\subheading{7. Photon interference with a {\caps QND}
path determination}
We describe next an  experimental scheme in which
a Mach-Zehnder interferometer is again used for measuring  an interference
\ob; but this time an additional component is added that is capable of carrying
out a nondemolishing path determination.
The experimental set-up is sketched out in Figure 2.

\line{}

\noindent
[{\it Figure 2. Mach-Zehnder interferometer with a Kerr medium.}]

\line{}

\noindent
The new component is a Kerr medium placed in the
second arm of the interferometer, which
will couple the $b$-mode field
with a single mode probe field according to the interaction
$$
U_K\ =\ I_1\otimes e^{-i\lambda (N_2\otimes N_3)}, \tag 61
$$
where $N_3 = c^*c$ is the number observable of the probe field.
This coupling will not change the number of photons of the interferometer
field, but it will affect the phase of the probe field.
Therefore, analysing the latter
with a phase sensitive detector $D_K$ yields information on the number of
photons in the second arm of the interferometer.
At the same time the detectors
$D_1$ and $D_2$ register the number of photons $N_1 = a^*a$ and $N_2 =
b^*b$ emerging from the second beam splitter, exhibiting thereby the
possible interference pattern. Such a scheme was  proposed
recently,$^{29}$ with the quadrature observable
$\frac 12(c^*+c)$ employed as the
readout \ob\ for the probe mode.
In fact any phase observable conjugate to $N_3$ would do as well.

Let $T$, $\kb 00$, and $T'$ be the input states of the incoming photon
pulse, the $b$-mode, and the probe field. One is again facing the task of
splitting the whole experiment into an observed and an observing part, or
into a preparation and a registration. Taken as a whole,
the state of the total
field will change in the Mach-Zehnder-Kerr apparatus according to
$$
T\otimes\kb 00\otimes T'\ \mapsto \
W\ :=\ U_\beta U_KV_{\delta}U_\alpha\,\bigl(T\otimes\kb 00\otimes T'\bigr)
\,U^*_\alpha V^*_{\delta} U^*_KU^*_\beta.\tag 62
$$
The probability of detecting $n$ photons in the counter $D_1$
and reading a value in a set $X$ in the homodyne detector $D_K$ is therefore
$$
p^{N_1\otimes I_2\otimes E}_W(n,1,X)\ =\
 \tr{W\,\kb nn\otimes I_2\otimes E(X)},\tag 63
$$
where $E$ is some (phase sensitive) readout \ob\ of the third mode
(such as the spectral measure of a quadrature component).
For simplicity, we have omitted now the detector $D_2$, since the
statistics of $D_1$ is already sufficient for indicating the
possible  interference  phenomenon.

The detection statistics (63) can again be interpreted
in various ways with respect to the input state of
an appropriate system.
We consider the incoming photon pulse, the $a$-mode, as the
input system. The  statistics (63) define then, for each
initial state $T'$ of the probe field, an observable
$A^{0,T'}$ associated with  the $a$-mode field such that
for all initial states $T$ of that
mode and for all $D_1$-values $n$
and $D_K$-values $X$,
$$
p^{A^{0,T'}}_T(n,X) :=
p^{N_1\otimes I_2\otimes E}_W(n,1,X). \tag 64
$$
This observable is an `enriched' version of the first marginal $F^{0;\eps}_1$,
Eq.\ (48), of the observable (47).
It depends on the whole variety of the ``apparatus parameters"
$\kb 00$, $T'$, $\alpha$, $\beta$, $\delta$, and $\lambda$.
Unlike $F^{0;\eps}_1$, the observable $A^{0,T'}$ contains direct
information on the dual aspects of a single photon. To see this, we
determine this observable by
specifying the beam splitters to be semitransparent
($\eps_1 = \eps_2 =\frac 12$) with $\vartheta_1 =
\vartheta_2 =\frac {\pi}2$, the case where one expects optimal interference.
The observable $A^{0,T'}$ is now found by  direct computation:
$$
\aligned
A^{0,T'}(n,X) \
=\ &\sum_{m=0}^\infty\textstyle{\frac{(m+n)!}{m!n!}}\,\kb{m+n}{m+n}\\
&\text{tr}\Bigl[T' \sin^n\bigl(\textstyle{\frac{\delta}2}I_3-
\textstyle{\frac{\lambda}2}N_3\bigr)\,
\cos^m\bigl(\textstyle{\frac{\delta}2}I_3-\textstyle{\frac{\lambda}2}N_3
\bigr)\,
e^{-i(m+n)\frac{\lambda}2N_3}\\
&E(X)\,
 e^{i(m+n)\frac{\lambda}2N_3}
\sin^n\bigl(\textstyle{\frac{\delta}2}I_3-\textstyle{\frac{\lambda}2}N_3\bigr)
\,
\cos^m\bigl(\textstyle{\frac{\delta}2}I_3-
\textstyle{\frac{\lambda}2}N_3\bigr)\Bigr].
\endaligned\tag 65
$$

One may go on to determine the marginal observables of  $A^{0,T'}$.
The first one is  obtained by putting $X = \bold R$:
$$
\aligned
A_1^{0,T'}(n)\ =\
\sum_{m=0}^\infty\textstyle{\frac{(m+n)!}{m!n!}}&\kb{m+n}{m+n}\\
\text{tr}\Bigl[T' &\sin^{2n}(\textstyle{\frac{\delta}2}I_3-
\textstyle{\frac{\lambda}2}N_3)
\cos^{2m}(\textstyle{\frac{\delta}2}I_3-\textstyle{\frac{\lambda}2}N_3)\Bigr],
\endaligned\tag 66
$$
showing that $A_1^{0,T'}$   is an unsharp number observable.
The second marginal observable
$X\mapsto A^{0,T'}_2(X) := \sum_n A^{0,T'}(n,X)$
is also directly obtained from (65) but it is less straightforward to
exhibit a simplified  expression for it.
However, this is not needed here since our primary interest is in
the case of the single photon input state $T = \kb 11$.
For this  state all the relevant probabilities are easily computed:
$$
\align
p^{A^{0,T'}}_{\ket 1}(0,X) &=
\text{tr}\Bigl[T'\cos\bigl(\textstyle{\frac{\delta}2}I_3-
\textstyle{\frac{\lambda}2}N_3\bigr)
e^{-i{\frac{\lambda}2}N_3}E(X)
e^{i{\frac{\lambda}2}N_3}
\cos\bigl(\textstyle{\frac{\delta}2}I_3-\textstyle{\frac{\lambda}2}N_3\bigr)
\Bigr],\ \ \ \ \ \ \tag 67a\\
p^{A^{0,T'}}_{\ket 1}(1,X) &=
\text{tr}\Bigl[T'\sin\bigl(\textstyle{\frac{\delta}2}I_3-
\textstyle{\frac{\lambda}2}N_3\bigr)
e^{-i{\frac{\lambda}2}N_3}E(X)
e^{i{\frac{\lambda}2}N_3}
\sin\bigl(\textstyle{\frac{\delta}2}I_3-\textstyle{\frac{\lambda}2}N_3\bigr)
\Bigr],\ \ \ \ \ \ \tag 67b\\
p^{A^{0,T'}}_{\ket 1}(0,\bold R)\, &=\,
p^{A^{0,T'}_{1}}_{\ket 1}(0)\, =\,
\text{tr}\Bigl[T'\cos^2\bigl(\textstyle{\frac{\delta}2}I_3-
\textstyle{\frac{\lambda}2}N_3\bigr)\Bigr],\tag 68a\\
p^{A^{0,T'}}_{\ket 1}(1,\bold R)\, &=\,
p^{A^{0,T'}_{1}}_{\ket 1}(1)\, =\,
\text{tr}\Bigl[T'\sin^2\bigl(\textstyle{\frac{\delta}2}I_3-
\textstyle{\frac{\lambda}2}N_3\bigr)\Bigr],\tag 68b\\
p^{A^{0,T'}}_{\ket 1}(\bold N,X)\, &=\,
p^{A^{0,T'}}_{\ket 1}(0,X) + p^{A^{0,T'}}_{\ket 1}(1,X)
\,=\, p^{A^{0,T'}_{2}}_{\ket 1}(X)\tag 69\\
&=\, \ts12\,\text{tr}\bigl[T'E(X)\bigr]\, +\,\ts12\,
\text{tr}\bigl[T'e^{-i\lambda N_3}\,E(X)\,e^{i\lambda N_3}\bigr].
\endalign
$$

As a consistency check one may first observe that for $\lambda = 0$
the probabilities (68) are just the single photon counting
probabilities in a Mach-Zehnder interferometer with $\eps_1 =
\eps_2 = \frac 12$ and $\vartheta_1 = \vartheta_2 =\frac{\pi}2$.
The introduction of the Kerr medium ($\lambda\ne 0$) in the second
arm of the interferometer affects these probabilities with a $T'$-dependent
phase shift.
Moreover, the detector $D_K$ allows one to collect now
an additional single-mode statistics (69) which contains information on
the path of the photon.
It must be emphasised that the $a$-mode observable $A^{0,T'}$  and the
ensuing single photon probability measures $p^{A^{0,T'}}_{\ket 1}$ do depend
on the state of the probe field $T'$ as well as on the path-indicating
observable $E$. Till now no properties of $T'$ or $E$ are used, and the
problem is to choose these
``control parameters" in such a way that the probabilities
(68-69) would display a good interference pattern together with
a reliable path determination.
Clearly, if  $T'$ is a number state or $E$ is compatible with the
number observable $N_3$ there will be no path information available from
(69). On the other hand,
if $E$ is any phase observable conjugate to the number observable $N_3$,
characterised by the covariance property
$e^{-i\lambda N_3}E(X)e^{i\lambda N_3}= E(X-\lambda)$,
then there is a possibility of obtaining information about
the path:
a photon traversing through the second arm of the interferometer leaves
a trace in the statistics of the second marginal $A^{0,T'}_2$
collected at the detector $D_K$.
In Ref.\ 29 $E$ was chosen to be the observable
$\frac 12 (c+c^*)$
and  $T'$  a coherent or squeezed state.

We  may again describe the whole experiment on the basis of the
`realistic' cut introduced in the preceding subsection, where
the first beam splitter and the phase shifter belong to the
preparation device. We assume from the outset that the $b$-mode
is idle so that the prepared state $S$ of the $(a,b)$-mode field is
$$
S\ :=\ V_\delta U_\alpha\,\bigl(T\otimes\kb 00\bigr)\,
     U^*_{\alpha} V^*_{\delta}.\tag 70
$$
From the counting statistics (63) one then obtains a unique
\ob\ $E^{T';\eps_2}$
of the $(a,b)$-mode field for any fixed input state $T'$ of the probe mode:
$$
p_S^{E^{T';\eps_2}}(n,X)\
=\ \text{tr}\left[S\otimes T'\,U_K^* U_{\beta}^*\,
    \bigl(\kb nn\otimes I_2\otimes E(X)\bigr)\,U_{\beta} U_K\right].\tag 71
$$
For the explicit determination of this \ob\ we shall consider only
a single photon
input state $T=\kb 11$ so that the object system is
given again by the subspace of states $\psi_{\alpha,\delta}$ from Eq.\
 (55).  On that state space the \ob\ $E^{T';\eps_2}$
is reduced to the following:
$$
F^{T';\eps_2}(n,X)\ :=\
\bigl(P_{10}+P_{01}\bigr)\,E^{T';\eps_2}(n,X)\,
\bigl(P_{10}+P_{01}\bigr).\tag 72
$$
Evaluation of Eq.\ (35) yields
$$
\align
E^{T';\eps_2}(n,X)\
&=\  \sum_{n_1,m_1,n_2,m_2}\kb{n_1,n_2}{m_1,m_2}\tag 73\\
   \big\langle{n_1,n_2}\big| &U_{\beta}^*\,\bigl(\kb nn\otimes
      I_2\bigr)\,U_\beta \big| m_1,m_2\big\rangle \,
\text{tr}\Bigl[T'\,e^{i\lambda m_1 N_3}\,E(X)\,e^{-i\lambda m_2 N_3}\Bigr].
\endalign
$$
The observable (72) is obtained by simply  carrying out the above
sum under the constraint $n_1+n_2=m_1+m_2=1$:
$$
\align
F^{T';\eps_2}(n,X)\ =\ \ &\kb{10}{10}\,
  \bigl[\eps_2 \delta_{n1}\,+\,(1-\eps_2 )\delta_{n0}\bigr]\,
  \text{tr}\Bigl[T'E(X)\Bigr]\tag 74\\
  +\ &\kb{01}{01}\,
  \bigl[(1-\eps_2) \delta_{n1}\,+\,\eps_2\delta_{n0}\bigr]\,
  \text{tr}\Bigl[T'e^{i\lambda N_3}E(X)e^{-i\lambda N_3}\Bigr]\\
  +\ &\kb{10}{01}\,\sqrt{\eps_2(1-\eps_2)}\,e^{i\vartheta_2}\,
    \bigl(\delta_{n1}-\delta_{n0}\bigr)\,
    \text{tr}\left[T'E(X)e^{-i\lambda N_3}\right]\\
  +\ &\kb{01}{10}\,\sqrt{\eps_2(1-\eps_2)}\,e^{-i\vartheta_2}\,
    \bigl(\delta_{n1}-\delta_{n0}\bigr)\,
    \text{tr}\left[T'e^{i\lambda N_3}E(X)\right].
\endalign
$$
It is instructive to determine the marginal observables
$F^{T';\eps_2}_1(n)$ :=
$F^{T';\eps_2}(n,\bo R)$ and
$F^{T';\eps_2}_2(X):=\sum_n F^{T';\eps_2}(n,X)$:
$$
\align
F^{T';\eps_2}_1(1)\ =\ &\ \kb{10}{10}\,\eps_2\,
        +\,\kb{01}{01}\,(1-\eps_2)\\
        &+\,\kb{10}{01}\,\sqrt{\eps_2(1-\eps_2)}\,e^{i\vartheta_2}\,
     \text{tr}\left[T'e^{-i\lambda N_3}\right]\tag 75a\\
&+\,\kb{01}{10}\,\sqrt{\eps_2(1-\eps_2)}\,e^{-i\vartheta_2}\,
         \text{tr}\left[T'e^{i\lambda N_3}\right],\\
F^{T';\eps_2}_1(0)\ =\ &\ \kb{10}{10}\,(1-\eps_2)\,
        +\,\kb{01}{01}\,\eps_2\\
        &+\,\kb{10}{01}\,\sqrt{\eps_2(1-\eps_2)}\,e^{i\vartheta_2}\,
     \text{tr}\left[T'e^{-i\lambda N_3}\right]\tag 75b\\
&+\,\kb{01}{10}\,\sqrt{\eps_2(1-\eps_2)}\,e^{-i\vartheta_2}\,
         \text{tr}\left[T'e^{i\lambda N_3}\right],\\
F^{T';\eps_2}_2(X)\ =\ &\ \kb{10}{10}\,\tr{T'E(X)}
       +\,\kb{01}{01}\,\text{tr}\left[T'\,e^{i\lambda N_3}E(X)e^{-i\lambda N_3}
\right].\tag 76\\
\endalign
$$
It is obvious that irrespective of the choice of $T'$ and the \ob\ $E$,
the first marginal is a smeared interference \ob, while
the second one is a smeared path \ob. This shows that the device serves
its purpose of establishing a joint \mt\ of these complementary properties.
With reference to the latter marginal it should be noted that neither $T'$
nor $E$ may commute with $N_3$ since otherwise the phase sensitivity is
lost. One may proceed to analyse Eqs.\ (75,\,76), showing that high path
confidence will lead to low interference contrast, and vice versa. Indeed,
an optimal interference would be obtained if $\eps_2=\frac 12$ and if
the numbers $\text{tr}\left[e^{\pm i\lambda N_3}T'\right]$ were equal to unity.
But this requires $T'$ to be an eigenstate of $N_3$, which destroys the
path \mt. On the other hand, imagine that  $E$ is a phase \ob,
$X=[0,\frac\pi 2]$, and $T'$ is chosen such that the number $\tr{T'E(X)}$
is close to unity. This can be achieved, for example,
for a coherent state with a
large amplitude. Such states have a slowly varying number distribution so
that the modulus of the complex numbers
$\text{tr}\left[e^{\pm i\lambda N_3}T'\right]$
is small compared to unity. But this is to say that the
first marginal \ob\ is also `close' to a path \ob, thus yielding only low
interference contrast.


There exists another scheme of a joint \mt\ of complementary path and
interference \ob{}s that is based solely on mirrors, beam splitters and
phase shifters, as they were used in the original Mach-Zehnder interferometer
(Figure 3).

\line{}

\noindent
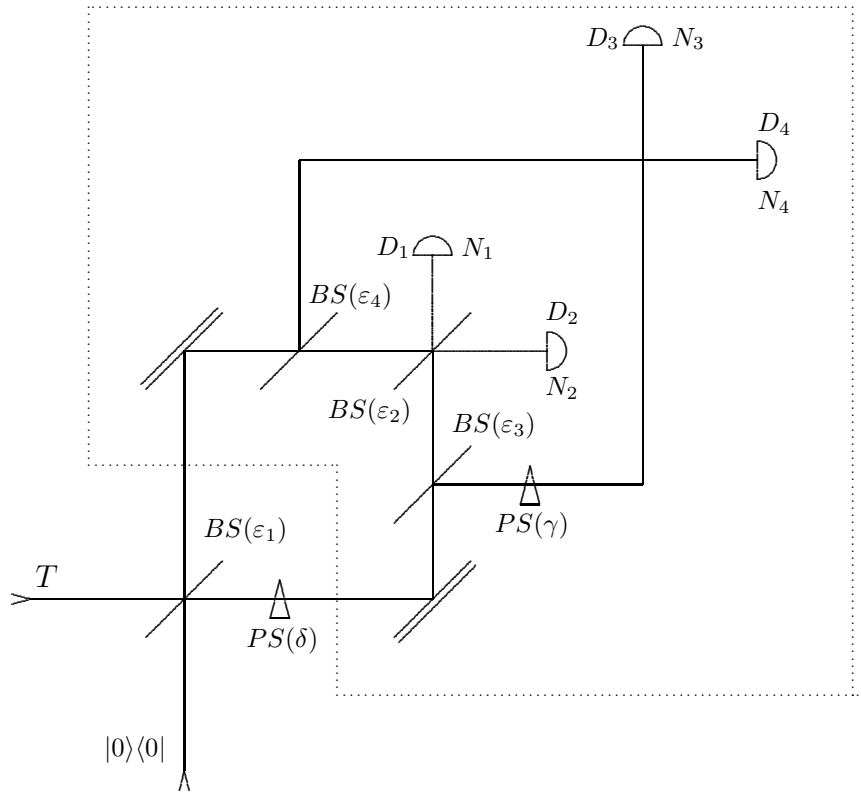
[{\it Figure 3. Expanded Mach-Zehnder interferometer}]

\line{}

\noindent
Regarding again the first beam splitter and the first phase shifter as
parts of the preparation device, the residual elements constitute the
measuring instrument which now has four detectors. An explicit analysis of
this experiment with respect to a single-photon input has been carried out
elsewhere,$^{30}$ so that here we may restrict ourselves to a
nontechnical summary.
The detection statistics again
give  rise to a unique observable of the object, which now has four
outcomes. One may combine pairs of these outcomes to add them up to three
two-valued marginal observables. By a suitable adjustment of the system
parameters it is possible to ensure that two of these \ob{}s are path and
interference \ob{}s; moreover, one may achieve the result
that the full detection
statistics uniquely determine the prepared state, that is, the \mt\ can be
managed to be informationally complete.
In particular, the statistics would
allow  one to find out the values of $\alpha$ and $\delta$.

\subheading{8. Conclusion}
In this paper we have elucidated the main measurement-theoretical
implications of the various versions of complementarity.
On the basis of an exact formalisation we have confirmed the common view
that complementarity in its strongest sense implies the noncoexistence of
the observables in question. Probabilistic complementarity, which is
equivalent to the \mt-theoretical version as long as only sharp observables
are involved, was found to persist for some pairs of unsharp \ob{}s even when
these are coexistent. In this way the apparent contradiction between
the possibility of joint \mt{}s of ``complementary" pairs of \obs\ is resolved
by taking into account unsharp \obs:
a complementary pair of sharp \obs\ can be replaced with a pair of smeared
versions which are then coexistent though still probabilistically \cpl{}y.

Starting from a quantum optical framework we have reconstructed the
common description of single-photon interferometer experiments in a
language which associates complementary path and interference \ob{}s
with the mutually exclusive options of detecting particle or wave phenomena.
Furthermore, it was found that introducing additional elements into the
Mach-Zehnder interferometer yields the possibility of jointly
measuring the path and observing interference patterns, in one single
experiment. The ensuing statistics is that of a joint \ob\ for a
coexistent pair of unsharp path and unsharp interference \ob{}s.
Bohr's original conception of complementarity as referring to a strict
mutual exclusiveness is thus confirmed for pairs of sharp \ob{}s.
Einstein's attempt to evade the complementarity verdict can be carried out
in some sense, but there is a price: lifting the \mt-theoretical
complementarity must be paid for by introducing a certain degree of
unsharpness, so that increasing confidence in the path determination
goes along with  decreasing interference contrast, and vice versa.

Finally it may be worth noting
that there is not just one unique way of giving a theoretical account
of one and the same experiment. What is regarded as part of physical
reality depends appreciably on the physicists' decision to place the
Heisenberg cut between preparation and measurement. In the present
example it becomes very obvious that quite different descriptions may arise.
The crucial point is that {\it some}  decision has to be made and that
each corresponds to adopting a certain {\it interpretation}. In our view
it seems  natural to adopt the most realistic picture which is
anyway used in practice as it serves as the best guide for physical
intuition.
\vskip 6pt

\subheading{References and footnotes}

\item{
1.\ } According to Wolfgang Pauli, the new quantum theory could have been
called the theory of complementarity. (W.\ Pauli, {\it General Principles
of Quantum Mechanics}, Springer, Berlin, 1980. Original German edition: 1933).
This is one example showing
the central importance of the notion of complementarity in
the discussions on the
foundations of quantum mechanics.

\item{
2.\ }  Ph.\ Grangier, G.\ Roger, A.\ Aspect.\
{\it Europhys.\ Lett.} {\bf 1} (1986) 173-179.
A.\ Aspect, Ph.\ Grangier.\
{\it Hyperfine Interactions} {\bf 37} (1987) 3-18.
A.\ Aspect, Ph.\ Gran\-gier.\
In: {\it Sixty-Two Years of Uncertainty:
Historical, Philosophical, and Physical Inquiries
into the Foundations of Quantum Mechanics.} Ed.~A.I.~Miller,
Plenum, New York, 1990, pp.~45-59.

\item{
3.\ }
This approach is discussed in a greater detail, for instance,
 in the monograph
 {\it Operational Quantum Physics}, {\bf LNP Vol.\ m31},
Springer, Berlin, 1995 (in the press),
by P.\ Busch, M.\ Grabowski, and P.\ Lahti.


\item{
4.\ } For a more detailed presentation of measurement theory
the reader may wish to consult the monograph {\it The Quantum Theory of
Measurement}, {\bf LNP Vol.\ m2}, Springer, Berlin, 1991 coauthored by P.\
Mittelstaedt and the present authors.
Second, revised edition, 1995 (in the press).\

\item{
5.\ } J.\ von Neumann, {\it Mathematische Grundlagen der Quantenmechanik},
Springer, Berlin, 1981.\ Original  edition: 1932.

\item{
6.\ } G.\ Ludwig, {\it Foundations of Quantum Mechanics},
Springer, Berlin, 1983.

\item{
7.\ } E.B.\ Davies and J.T.\ Lewis, {\it Commun.\ math.\ Phys.} {\bf 17} (1970)
239-260.\
See also P.\ Lahti, in {\it Recent Development in Quantum Logic},
eds.\ P.\ Mittelstaedt and E.-W.\ Stachow, Bibliographisches Institut
Mannheim, 1985, pp 61-80.

\item{
8.\ } N.\ Bohr, {\it Nature} {\bf } (1928) 580-590;
{\it Phys.\ Rev.} {\bf 48} (1935) 696-702.

\item{
9.\ } We recall that for two bounded \sad\ operators $A$ and $B$ the
ordering $A\le B$ is defined as $\ip\fii{A\fii}\le \ip\fii{B\fii}$ for all
$\fii\in\hi$.

\item{
10.} For two bounded positive operators $A,B$ the greatest lower bound may
or may not exist among the positive operators. If it does exist it is
denoted $A\land B$.

\item{
11.} 
K.\ Kraus, {\it Phys.\ Rev.\ D} {\bf 10} (1987) 3070-3075.

\item{
12.} For pairs of state transformations $\Phi_1,\Phi_2$ the ordering
$\Phi_1\le \Phi_2$ is defined as $\Phi_1(T)\le \Phi_2(T)$ for all states
$T$. If for two state transformations $\Phi_1,\Phi_2$ the greatest lower
bound exists it is denoted $\Phi_1\land\Phi_2$.

\item{
13.} P.\ Busch and P.\ Lahti, {\it Phys.\ Rev.\ D} {\bf 29} (1984) 1634-1646.

\item{
14.} J.C.\ Garrison and J.\ Wong,
{\it J.\ Math.\ Phys.} {\bf 11} (1970) 2242-2249.

\item{
15.} A.\ Lenard,
{\it  J.\ Funct.\ Anal.} {\bf  10} (1972)  410-423.

\item{
16.} P.\ Lahti and K.\ Ylinen, {\it J.\ Math.\ Phys.} {\bf 28} (1987)
2614-2617.

\item{
17.} P.\ Busch and P.\ Lahti, {\it Phys.\ Lett.\ A} {\bf 115} (1986) 259 -264.

\item{
18.} K.\ Ylinen,
{\it J.\ Math.\ Anal.\ Appl.} {\bf  137} (1989) 185-192.

\item{
19.} W.\ Heisenberg, {\it Z.\ Phys.}   {\bf  43} (1927) 172-198.

\item{
20.} P.\ Busch, {\it Int.\ J.\ Theor.\ Phys.} {\bf 24} (1985) 63-92.

\item{
21.} E.B.\ Davies, {\it Quantum Theory of Open Systems}, Academic Press,
New York, 1976.
E.\ Prugove\v{c}ki, {\it Stochastic Quantum Mechanics and Quantum Spacetime},
Reidel, Dordrecht, 1986.

\item{
22.} P.\ Busch, {\it Phys.\ Rev.\ D} {\bf 33}   (1986) 2253-2261.

\item{
23.} W.\ Mlak, {\it Hilbert Spaces and Operator Theory},
Kluwer, Dordrecht, 1991.


\item{
24.} P.\ Busch, {\it J.\ Math.\ Phys.} {\bf 25} (1984) 1794-1797.

\item{
25.} P.\ Busch, M.\ Grabowski and P.\ Lahti, {\it Ann.\ Phys.\ (N.Y.)}
{\bf 237} (1995) 1-11.

\item{
26.}    S.\ Prasad, M.O.\ Scully, W.\ Martienssen.
               {\it Opt.\ Commun.} {\bf 62} (1987) 139-145.
        R.A.\ Campos, B.E.A.\ Saleh, M.C.\ Teich.\
               {\it Phys.\ Rev.\ A} {\bf 40} (1989) 1371-1384.

\item{
27.}  A.M.\ Perelomov, {\it Generalized Coherent States
               and Their Applications},
               Sprin\-ger, Berlin, 1986.

\item{
28.}  P.\ Mittelstaedt, A.\ Prieur, R.\ Schieder.
{\it Found.\ Phys.} {\bf 17}
               (1987) 893-903.

\item{
29.}   B.C.\ Sanders, G.J.\ Milburn.\
{\it Phys.\ Rev.\ A} {\bf 39}
               (1989) 694-702.

\item{
30.}   P.\ Busch.\
{\it Found.\ Phys.} {\bf 17}
               (1987) 905-937.

\bye

\vfill\eject

\subheading{Figure Captions}

\line{}

\noindent
{\it Figure 1.\ Scheme of a Mach-Zehnder interferometer.}

\line{}

\noindent
{\it Figure 2.\ Mach-Zehnder interferometer with a Kerr medium.}

\line{}

\noindent
{\it Figure 3.\ Expanded Mach-Zehnder interferometer}

\bye